\documentclass{mn2e} 
\usepackage{color}
\usepackage{dcolumn}
\usepackage{bm}
\usepackage{epsfig}

\def\reference{\par\noindent\hangindent\parindent}
\def\kms{\,{\rm km}\,{\rm s}^{-1}}
\def\mpcoh{{\,h^{-1}\,\rm Mpc}}
\def\kpcoh{{\,h^{-1}\,\rm kpc}}
\font\tweak = cmr12 at 13 truept

\begin{document}

\twocolumn

\title[GAMA --- {\bf DRAFT v3.1}]{Galaxy and Mass Assembly (GAMA): survey diagnostics and core data release}

\author[Driver et al. --- {\bf DRAFT v3.1}] 
{\tweak S.P.Driver$^1$\thanks{SUPA, Scottish Universities Physics Alliance},
D.T.Hill$^1$,
L.S.Kelvin$^1$, 
A.S.G.Robotham$^1$, 
J.Liske$^2$, 
P.Norberg$^3$, 
I.K.Baldry$^4$, \newauthor \tweak
S.P.Bamford$^5$,
A.M.Hopkins$^6$,
J.Loveday$^7$, 
J.A.Peacock$^3$,
E.Andrae$^{8}$, 
J.Bland-Hawthorn$^9$, \newauthor \tweak
S.Brough$^6$, 
M.J.I.Brown$^{10}$,
E.Cameron$^{11}$,
J.H.Y.Ching$^9$,
M.Colless$^{6}$,
C.J.Conselice$^5$, \newauthor \tweak 
S.M.Croom$^9$, 
N.J.G.Cross$^3$,
R.De Propris$^{12}$, 
S.Dye$^{13}$
M.J.Drinkwater$^{14}$,
S.Ellis$^9$, \newauthor \tweak 
Alister W.Graham$^{15}$, 
M.W.Grootes$^{15}$,
M.Gunawardhana$^{8}$, 
D.H.Jones$^6$,
E.van Kampen$^2$, \newauthor \tweak 
C.Maraston$^{12}$, 
R.C.Nichol$^{16}$, 
H.R.Parkinson$^3$, 
S.Phillipps$^{17}$, 
K.Pimbblet$^{10}$,
C.C.Popescu$^{18}$, \newauthor \tweak 
M.Prescott$^4$,   
I.G.Roseboom$^7$,
E.M.Sadler$^9$, 
A.E.Sansom$^{18}$, 
R.G.Sharp$^6$,   
D.J.B.Smith$^5$, \newauthor \tweak 
E.Taylor$^{8,20}$, 
D.Thomas$^{16}$, 
R.J.Tuffs$^{8}$,  
D.Wijesinghe$^{9}$,
L.Dunne$^5$,   
C.S.Frenk$^{22}$,
M.J.Jarvis$^{23}$, \newauthor \tweak   
B.F.Madore$^{24}$, 
M.J.Meyer$^{25}$, 
M.Seibert$^{24}$, 
L.Staveley-Smith$^{25}$, 
W.J.Sutherland$^{26}$, 
S.J.Warren$^{27}$\\ \\
$^1$ School of Physics \& Astronomy, University of St Andrews, North Haugh, St Andrews, KY16 9SS, UK; SUPA\\
$^2$ European Southern Observatory, Karl-Schwarzschild-Str. 2, 85748, Garching, Germany\\
$^3$ Institute for Astronomy, University of Edinburgh, Royal Observatory, Blackford Hill, Edinburgh, EH9 3HJ, UK\\
$^4$ Astrophysics Research Institute, Liverpool John Moores University, Twelve Quays House, Egerton Wharf, Birkenhead, CH4 1LD, UK\\
$^5$ Centre for Astronomy and Particle Theory, University of Nottingham, University Park, Nottingham, NG7 2RD, UK\\
$^6$ Australian Astronomical Observatory, PO Box 296, Epping, NSW 1710, Australia\\
$^7$ Astronomy Centre, University of Sussex, Falmer, Brighton, BN1 9QH, UK\\
$^{8}$ Max-Plank Institute for Nuclear Physics (MPIK), Saupfercheckweg 1, 69117 Heidelberg, Germany\\
$^9$ Sydney Institute for Astronomy, School of Physics, University of Sydney, NSW 2006, Australia\\
$^{10}$ School of Physics, Monash University, Clayton, Victoria 3800, Australia\\
$^{11}$ Department of Physics, Swiss Federal Institute of Technology (ETH-Z\"urich), 8093 Z\"urich, Switzerland\\
$^{12}$ Cerro Tololo Inter-American Observatory, La Serena, Chile\\
$^{13}$ School of Physics and Astronomy, Cardiff University, Queens Buildings, The Parade, Cardiff, CF24 3AA, UK\\
$^{14}$ Department of Physics, University of Queensland, Brisbane, Queensland 4072, Australia \\
$^{15}$ Centre for Astrophysics and Supercomputing, Swinburne University of Technology, Hawthorn, Victoria 3122, Australia\\
$^{16}$ Institute of Cosmology and Gravitation (ICG), University of Portsmouth, Dennis Sciama Building Road, Portsmouth, PO1 3FX, UK\\
$^{17}$ Astrophysics Group, H.H. Wills Physics Laboratory, University of Bristol, Tyndall Avenue, Bristol BS8 1TL, UK\\
$^{18}$ Jeremiah Horrocks Institute, University of Central Lancashire, Preston, PR1 2HE, UK\\
$^{19}$ Instituto Astronomico e Geoffisico, University of San Paulo, Brazil\\
$^{20}$ School of Physics, University of Melbourne, Victoria 3010, Australia\\
$^{21}$ Institute for Computational Cosmology, Department of Physics, University of Durham South Road, Durham, DH1 3LE, UK \\
$^{22}$ Centre for Astrophysics, Science \& Technology Research Institute, University of Hertfordshire, Hatfield, AL10 9AB, UK\\
$^{23}$ Observatories of the Carnegie Institute of Washington, 813 Santa Barbara Street, Pasadena, CA91101, USA\\
$^{24}$ International Centre for Radio Astronomy Research, The University of Western Australia, 35 Stirling Hwy, Crawley, WA 6009, Australia\\
$^{25}$ Astronomy Unit, Queen Mary University London, Mile End Road, London, E1 4NS, UK\\
$^{26}$ Astrophysics Group, Imperial College London, Blackett Laboratory, Prince Consort Road, London, SW7 2AZ, UK}
\pubyear{2010} \volume{000}
\pagerange{\pageref{firstpage}--\pageref{lastpage}}

\maketitle
\label{firstpage}

\begin{abstract}
The Galaxy And Mass Assembly (GAMA) survey has been operating since
February 2008 on the 3.9-m Anglo-Australian Telescope using the
AAOmega fibre-fed spectrograph facility to acquire spectra with a
resolution of $R\approx1300$ for 120\,862 SDSS selected galaxies. The
target catalogue constitutes three contiguous equatorial regions
centred at 9$^h$ (G09), 12$^h$ (G12) and 14.5$^h$ (G15) each of $12
\times 4$ deg$^2$ to limiting fluxes of $r_{\rm pet} < 19.4$, $r_{\rm
  pet} < 19.8$, and $r_{\rm pet} < 19.4$ mag respectively (and
additional limits at other wavelengths). Spectra and reliable
redshifts have been acquired for over 98 per cent of the galaxies
within these limits. Here we present the survey footprint,
progression, data reduction, redshifting, re-redshifting, an
assessment of data quality after 3 years, additional image analysis
products (including $ugrizYJHK$ photometry, S\'ersic profiles and
photometric redshifts), observing mask, and construction of our core
survey catalogue ({\sf GamaCore}). From this we create three science
ready catalogues: {\sf GamaCoreDR1} for public release, which includes
data acquired during year 1 of operations within specified magnitude
limits (February 2008 to April 2008); {\sf GamaCoreMainSurvey}
containing all data above our survey limits for use by the GAMA team
and collaborators; and {\sf GamaCoreAtlasSv} containing year 1, 2 and
3 data matched to Herschel-ATLAS Science Demonstration data. These
catalogues along with the associated spectra, stamps, and profiles,
can be accessed via the GAMA website: {\tt
  http://www.gama-survey.org/}
\end{abstract}

\setlength{\extrarowheight}{0pt}

\section{Introduction}
Large scale surveys are now a familiar part of the astronomy landscape
and assist in facilitating a wide range of science programmes. Three of
the most notable wide-area surveys in recent times, with a focus on
galactic and galaxy evolution, are the 2 Micron All Sky Survey (2MASS;
Skrutskie et al.~2006), the 2 degree Field Galaxy Redshift Survey
(2dFGRS; Colless et al.~2001; 2003), and the Sloan Digital Sky Survey
(SDSS; York et al.~2000). These surveys have each been responsible for
a wide range of science advances attested by their publication and
citation records (Trimble \& Ceja 2010), ranging from: the
identification of new stellar types (e.g., Kirkpatrick et al., 1999);
tidal streams in the Galactic Halo (e.g., Belokurov et al.~2006); new
populations of dwarf galaxies (e.g., Willman et al.~2005); galaxy
population statistics (e.g., Bell et al.~2003; Baldry et al.~2006);
the recent cosmic star-formation history (e.g., Heavens et al.~2004);
group catalogues (e.g., Eke et al.~2004); merger rates (e.g., Bell et
al., 2006); quantification of large scale structure (e.g., Percival et
al.~2001); galaxy clustering (e.g., Norberg et al.~2001); and, in
conjunction with CMB and SNIa searches, convergence towards the basic
cosmological model now adopted as standard (e.g., Spergel et al.~2003;
Cole et al.~2005). In addition to these mega-surveys there have been a
series of smaller, more specialised, local surveys including the
Millennium Galaxy Catalogue (MGC; Driver et al., 2005), the 6dF Galaxy
Survey (6dFGS; Jones et al.~2004, 2009), the HI Parkes All Sky Survey
(HIPASS; Meyer et al.~2004), and the Galaxy Evolution Explorer (GALEX;
Martin et al.~2005) mission, which are each opening up new avenues of
extra-galactic exploration (i.e., structural properties, the near-IR
domain, the 21cm domain, and the UV domain respectively). Together
these surveys provide an inhomogeneous nearby reference point for the
very narrow high-z pencil beam surveys underway (i.e., DEEP2, VVDS,
COSMOS, GEMS etc.), and from which comparative studies can be made to
quantify the process of galaxy evolution (e.g., Cameron \& Driver
2007).

The Galaxy And Mass Assembly Survey (GAMA) has been established with
two main sets of aims, which are surveyed in Driver et al.~(2009). The
first is to use the galaxy distribution to conduct a series of tests
of the Cold Dark Matter (CDM) paradigm, and the second is to carry out
detailed studies of the internal structure and evolution of the
galaxies themselves.

The CDM model is now the standard means by which data relevant to
galaxy formation and evolution are interpreted, and it has met with
great success on 10--100 Mpc scales. The next challenge in validating
this standard model is to move beyond small linear fluctuations, into
the regime dominated by dark-matter haloes. The fragmentation of the
dark matter into these roughly spherical virialized objects is
robustly predicted both numerically and analytically over seven orders
of magnitude in halo mass (e.g., Springel et al. 2005). Massive haloes
are readily identified as rich clusters of galaxies, but it remains a
challenge to probe further down the mass function. For this purpose,
one needs to identify low-mass groups of galaxies, requiring a survey
that probes far down the galaxy luminosity function over a large
representative volume. But having found low-mass haloes, the galaxy
population within each halo depends critically on the interaction
between the baryon processes (i.e., star formation rate and feedback
efficiency) and the total halo mass. In fact the ratio of stellar mass
to halo mass is predicted (Bower et al.~2006; De Lucia et al.~2006)
to be strongly dependent on halo mass, exhibiting a characteristic dip
at Local Group masses. The need for feedback mechanisms to suppress
star formation in both low mass haloes (via supernovae) and high mass
haloes (via AGN) is now part of standard prescriptions in modeling
galaxy formation. With GAMA we can connect these theoretical
ingredients directly with observational measurements.

However, the astrophysics of galaxy bias is not the only poorly
understood area in the CDM model. Existing successes have been
achieved at the price of introducing Dark Energy as the majority
constituent of the Universe, and a key task for cosmology is to
discriminate between various explanations for this phenomenon: a
cosmological constant, time varying scalar field, or a deficiency in
our gravity model. These aspects can be probed by GAMA in two distinct
ways: either the form or evolution of the halo mass function may
diverge from standard predictions of gravitational collapse in the
highly nonlinear regime, or information on non-standard models may be
obtained from velocity fields on 10-Mpc scales. The latter induce
redshift-space anisotropies in the clustering pattern, which measure
the growth rate of cosmic structure (e.g., Guzzo et al. 2008).  Thus
GAMA has the potential to illuminate both the astrophysical and
fundamental aspects of the CDM model.

Moving beyond the large-scale distribution, GAMA's main long-term
legacy will be to create a uniform galaxy database, which builds on
earlier local surveys in a comprehensive manner to fainter flux
levels, higher redshift, higher spatial resolution, and spanning UV to
radio wavelengths. The need for a combined homogeneous,
multiwavelength, and spatially resolved study can be highlighted by
three topical issues:


{\sf 1. Galaxy Structure.} Galaxies are typically comprised of bulge
and/or disc components that exhibit distinct properties (dynamics,
ages, metallicities, profiles, dust and gas content), indicating
potentially distinct evolutionary paths (e.g., cold smooth and hot
lumpy accretion; cf. Driver et al.~2006 or Cook et al.~2010). This is
corroborated by the existence of the many SMBH-bulge relations (see,
for example, Novak, Faber \& Dekel 2006), which firmly couples
spheroid-only evolution with AGN history (Hopkins et al.~2006). A
comprehensive insight into galaxy formation and evolution therefore
demands consideration of the structural components requiring high
spatial resolution imaging on $\sim 1$~kpc scales or better (e.g.,
Allen et al.~2006; Gadotti 2009).

{\sf 2. Dust attenuation.} A recent spate of papers (Shao et al.~2007;
Choi et al.~2007; Driver et al.~2007, 2008; Masters et al.~2010) have
highlighted the severe impact of dust attenuation on the measurement
of basic galaxy properties (e.g., fluxes and sizes). In particular,
dust attenuation is highly dependent on wavelength, inclination,
and galaxy type with the possibility of some further dependence on
environment. Constructing detailed models for the attenuation of
stellar light by dust in galaxies and subsequent re-emission (e.g.,
Popescu et al.~2000), is intractable without extensive wavelength
coverage extending from the UV through to the far-IR. To survey the
dust content for a significant sample of galaxies therefore demands a
multi-wavelength dataset extending over a sufficiently large volume to
span all environments and galaxy types. The GAMA regions are or will
be surveyed by the broader GALEX Medium Imaging Survey and
Herschel-ATLAS (Eales et al.~2010) programmes, providing UV to FIR
coverage for a significant fraction of our survey area.

{\sf 3. The HI content.} As star-formation is ultimately driven by a
galaxy's HI content any model of galaxy formation and evolution must
be consistent with the observed HI properties (see for example
discussion in Hopkins, McClure-Griffiths \& Gaensler 2008). Until
recently probing HI beyond very low redshifts has been laborious if
not impossible due to ground-based interference and/or sensitivity
limitations (see for example Lah et al.~2009). The new generation of
radio arrays and receivers are using radio-frequency interference
mitigation methods coupled with new technology receivers to open up
the HI Universe at all redshifts (i.e., ASKAP, MeerKAT, LOFAR, and
ultimately the SKA). This will enable coherent radio surveys that are
well matched in terms of sensitivity and resolution to optical/near-IR
data. Initial design study investment has been made in the DINGO
project, which aims to conduct deep HI observations within a
significant fraction of the GAMA regions using ASKAP (Johnston et
al.~2007).

The GAMA survey will eventually provide a wide-area highly complete
spectroscopic survey of over 400k galaxies with sub-arcsecond
optical/near-IR imaging (SDSS/UKIDSS/VST/VISTA). Complementary
multi-wavelength photometry from the UV (GALEX), mid-IR (WISE) and
far-IR (Herschel) and radio wavelengths (ASKAP, GMRT) is being
obtained by a number of independent public and private survey
programmes. These additional data will ultimately be ingested into the
GAMA database as they becomes available to the GAMA team. At the heart
of the survey is the 3.9-m Anglo Australian Telescope (AAT), which is
being used to provide the vital distance information for all galaxies
above well specified flux, size and isophotal detection limits. In
addition the spectra from AAOmega for many of the sample will be of
sufficiently high signal-to-noise and spectral resolution ($\approx 3$
--- $6\,$\AA) to allow for the extraction of line diagnostic information
leading to constraints on star-formation rates, velocity dispersions
and other formation/evolutionary markers. The area and depth of GAMA
compared to other notable surveys is detailed in Baldry et al. (2010;
their Fig.~1).  In general GAMA lies in the parameter space between
that occupied by the very wide shallow surveys and the deep pencil
beam surveys and is optimised to study structure on $\sim
10\mpcoh$ to $1\kpcoh$ scales, as well as sample the galaxy
population from FUV to radio wavelengths.

This paper describes the first three years of the GAMA AAOmega
spectroscopic campaign, which has resulted in $112$k new redshifts (in
addition to the $19$k already known in these regions). In section 2 we
describe the spectroscopic progress, data reduction, redshifting,
re-redshifting, an assessment of the redshift accuracy and blunder
rate, and an update to our initial visual classifications. In section
3 we describe additional image analysis resulting in $ugrizYJHK$
matched aperture photometry, S\'ersic profiles, and photometric
redshifts. In section 4 we describe the combination of the data
presented in section 2 and 3 to form our core catalogue and
investigate the completeness versus magnitude, colour, surface
brightness, concentration, and close pairs. In section 5 we present
the survey masks required for spatial clustering studies and in
section 6 present our publicly available science ready
catalogue. These catalogues along with an MYSQL tool and other data
inspection tools are now available at: {\tt
  http://www.gama-survey.org/} and we expect future releases of
redshifts and other data products to occur on an approximately annual
cycle.

Please note all magnitudes used in this paper, unless otherwise
specified, are $r$-band Petrosian ($r_{\rm pet}$) from SDSS DR6, which
have been extinction corrected and placed onto the true AB scale
following the prescription described by the SDSS DR6 release
(Adelman-McCarthy et al.~2008).

\section{The GAMA AAT spectroscopic survey}

\subsection{GAMA field selection, input catalogue and tiling algorithm}
\begin{figure*}
\centerline{\psfig{file=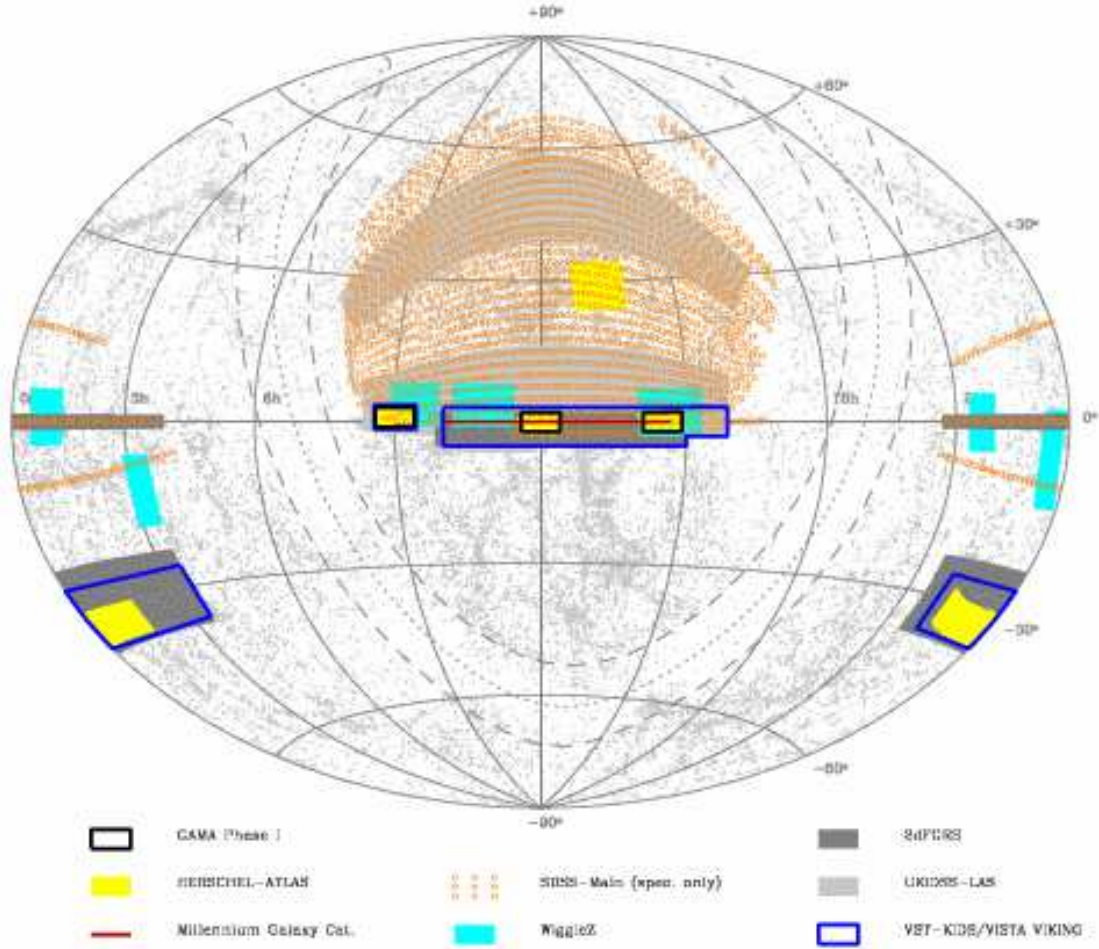,width=\textwidth}}

\vspace{-3.0cm}

\caption{GAMA Phase I (black squares) in relation to other recent and
  planned surveys (see key). For a zoom in to the GAMA regions showing
  SDSS, UKIDSS and GALEX overlap please see Fig.~1 of the companion
  paper describing the photometry by Hill et al., (2010a). Also
  overlaid as grey dots are all known redshifts at $z<0.1$ taken from
  NASA ExtraGalactic Database. \label{fig:scope}}
\end{figure*}

The initial GAMA survey consists of three equatorial regions, each of
12 $\times$ 4 deg$^2$ (see Table~\ref{tab:fields} and
Fig.~\ref{fig:scope}). The decision for this configuration was driven
by three considerations: suitability for large-scale structure studies
demanding contiguous regions of $\sim 50$ deg$^2$ to fully sample
$\sim100$ co-moving $\mpcoh$ structures at z $\approx$ 0.2;
observability demanding a 5~hr Right Ascension baseline to fill a
night's worth of observations over several lunations; and overlap with
existing and planned surveys, in particular the SDSS (York et
al. 2000), UKIDSS LAS (Lawrence et al. 2007), VST KIDS, VISTA VIKING,
H-ATLAS (Eales et al. 2010), and ASKAP DINGO. Fig.~\ref{fig:scope}
shows the overlap of some of these surveys. Note the H-ATLAS SGP
survey region has changed since Fig.~5 of Driver et al. (2009) due to
additional space-craft limitations introduced in-flight. The depth and
area of the GAMA AAT spectroscopic survey were optimised following
detailed simulations of the GAMA primary science goal of measuring the
Halo Mass Function. This resulted in an initial survey area of
144~deg$^2$ to a depth of $r_{\rm pet} < 19.4$ mag in the 9$^h$ and
15$^h$ regions and an increased depth of $r_{\rm pet} < 19.8$ mag in
the 12$^h$ region (see Table~\ref{tab:fields}). In addition for year 2
and 3 observations additional $K$ and $z$-band selection was
introduced such that the final main galaxy sample (Main Survey) can be
defined (see Baldry et al., 2010) as follows:

~

\noindent
{\bf G09:} $r_{\rm pet}<19.4$ {\sf OR} ($K_{\rm Kron} < 17.6$ {\sf AND} $r_{\rm model} <
20.5$) {\sf OR} ($z_{\rm model} < 18.2$ {\sf AND} $r_{\rm model} < 20.5$) mag

~

\noindent
{\bf G12:} $r_{\rm pet}<19.8$ {\sf OR} ($K_{\rm Kron} < 17.6$ {\sf AND} $r_{\rm model} < 20.5$) {\sf OR} ($z_{\rm model} < 18.2$ {\sf AND} $r_{\rm model} < 20.5$) mag 

~

\noindent
{\bf G15:} $r_{\rm pet}<19.4$ {\sf OR} ($K_{\rm Kron} < 17.6$ {\sf AND} $r_{\rm model} < 20.5$) {\sf OR} ($z_{\rm model} < 18.2$ {\sf AND} $r_{\rm model} < 20.5$) mag 

~

\noindent
with all magnitudes expressed in AB. For the remainder of this paper
we mainly focus, for clarity, on the $r$-band selected data and note
that equivalent diagnostic plots to those shown later in this paper
can easily be created for the $z$ and $K$ selections.

\begin{table*}
\caption{Coordinates of the three GAMA equatorial fields. \label{tab:fields}}
\begin{center}
\begin{tabular}{ccccc} \hline
Field & RA(J2000 deg.) & Dec(J2000 deg.) & Area (deg$^2$) & Depth (mag) \\ \hline 
G09 & $129.0 ... 141.0$ & $+3.0 ... -1.0$ & $12\times4$ & $r_{\rm pet} < 19.4$\\
G12 & $174.0 ... 186.0$ & $+2.0 ... -2.0$ & $12\times4$ & $r_{\rm pet} < 19.8$ \\
G15 & $211.5 ... 223.5$ & $+2.0 ... -2.0$ & $12\times4$ & $r_{\rm pet} < 19.4$ \\ \hline
\end{tabular}
\label{tab:gama-fields}
\end{center}
\end{table*}

\subsection{Survey preparation}
The input catalogue for the GAMA spectroscopic survey was constructed
from SDSS DR6 (Adelman-McCarthy et al. 2008) and our own reanalysis of
UKIDSS LAS DR4 (Lawrence et al. 2007) to assist in star-galaxy
separation; see Hill et al. (2010a) for details. The preparation of
the input catalogue, including extensive visual checks and the
revised star-galaxy separation algorithm, is described and assessed in
detail by Baldry et al. (2010). The survey is being conducted using
the AAT's AAOmega spectrograph system (an upgrade of the original 2dF
spectrographs: Lewis et al. 2002; Sharp et al. 2006).

In year 1 we implemented a uniform grid tiling algorithm (see
Fig.~\ref{fig:tiles}, upper panels), which created a significant
imprint of the tile positions on the spatial completeness
distribution. Subsequently, for years 2 and 3 we implemented a
heuristic ``greedy'' tiling strategy, which was designed to maximise
the spatial completeness across the survey regions within $0.14\deg$
smoothed regions. Full details of the GAMA science requirements and
the tiling strategy devised to meet these are laid out by Robotham et
al.~(2010). The efficiency of this strategy is discussed further in
section~\ref{sec:masks}. The final location of all tiles is shown in
Fig.~\ref{fig:tiles} and the location of objects for which redshifts
were not secured or not observed are shown in the lower panels of
Fig.~\ref{fig:dist} as black dots or red crosses respectively.

\subsection{Observations and data reduction}
\begin{figure*}

\hspace{-10.5cm}{\psfig{file=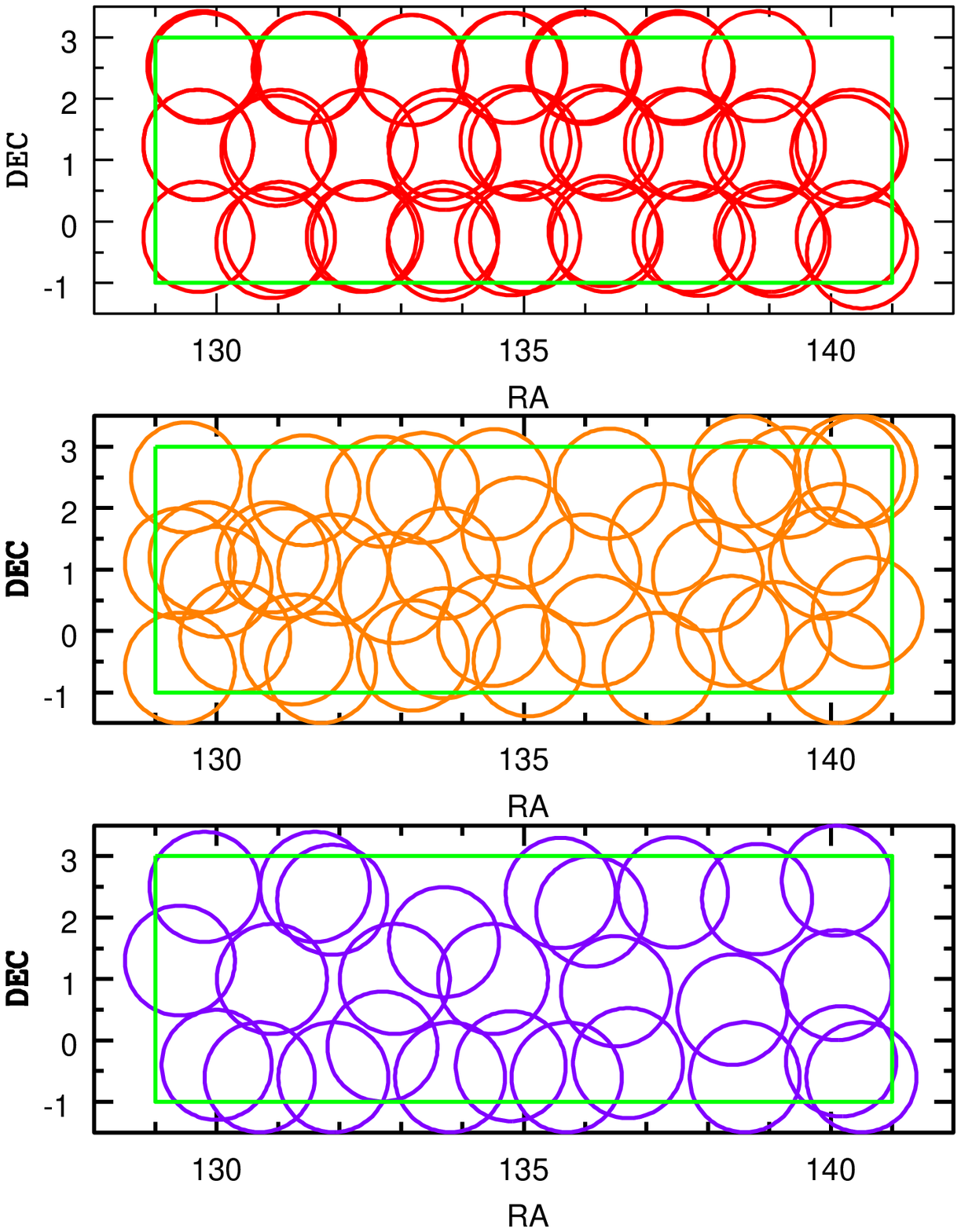,width=7.0cm}}

\vspace{-7.03cm}

\centerline{\psfig{file=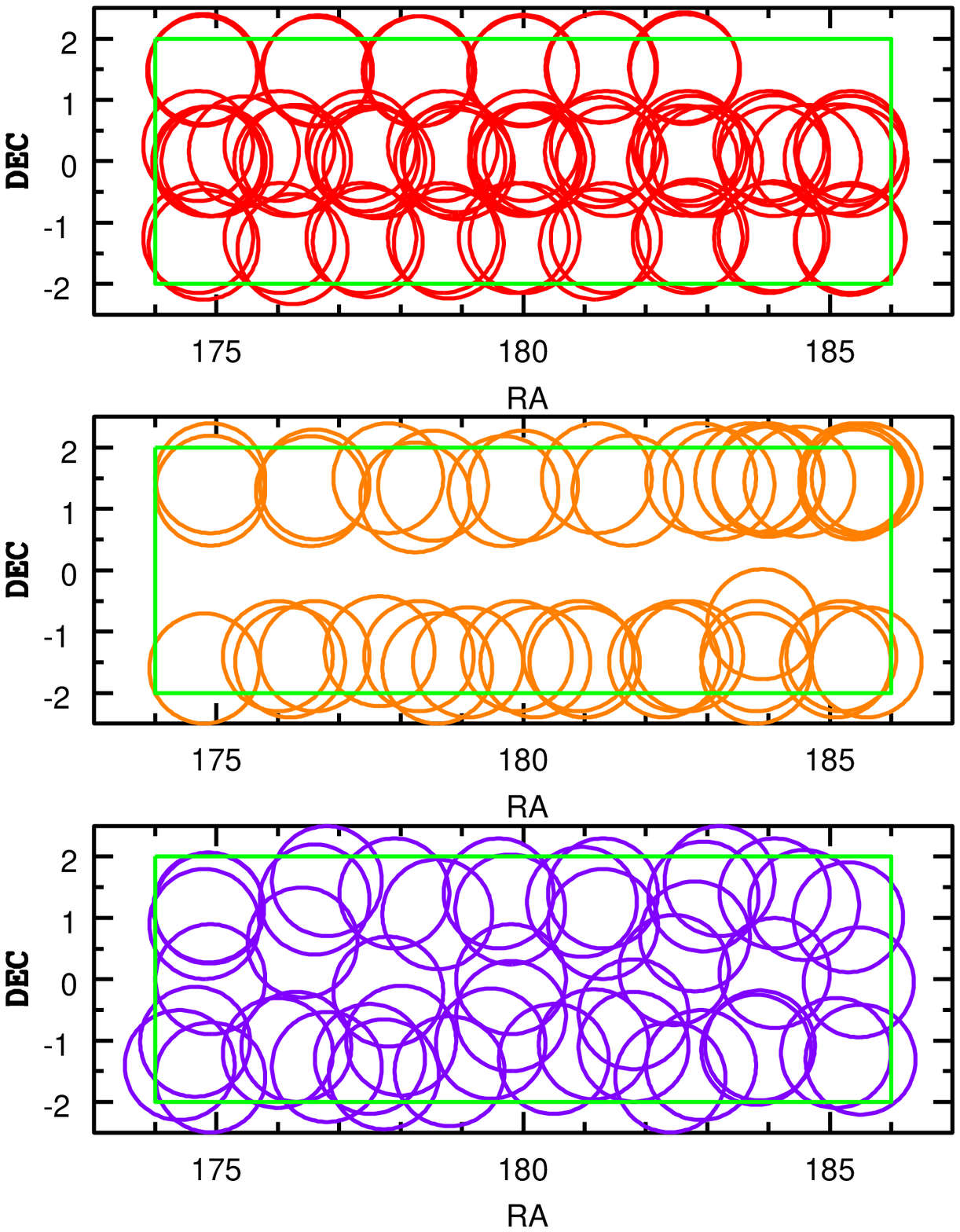,width=7.0cm}}

\vspace{-7.03cm}

\hspace{10.5cm}{\psfig{file=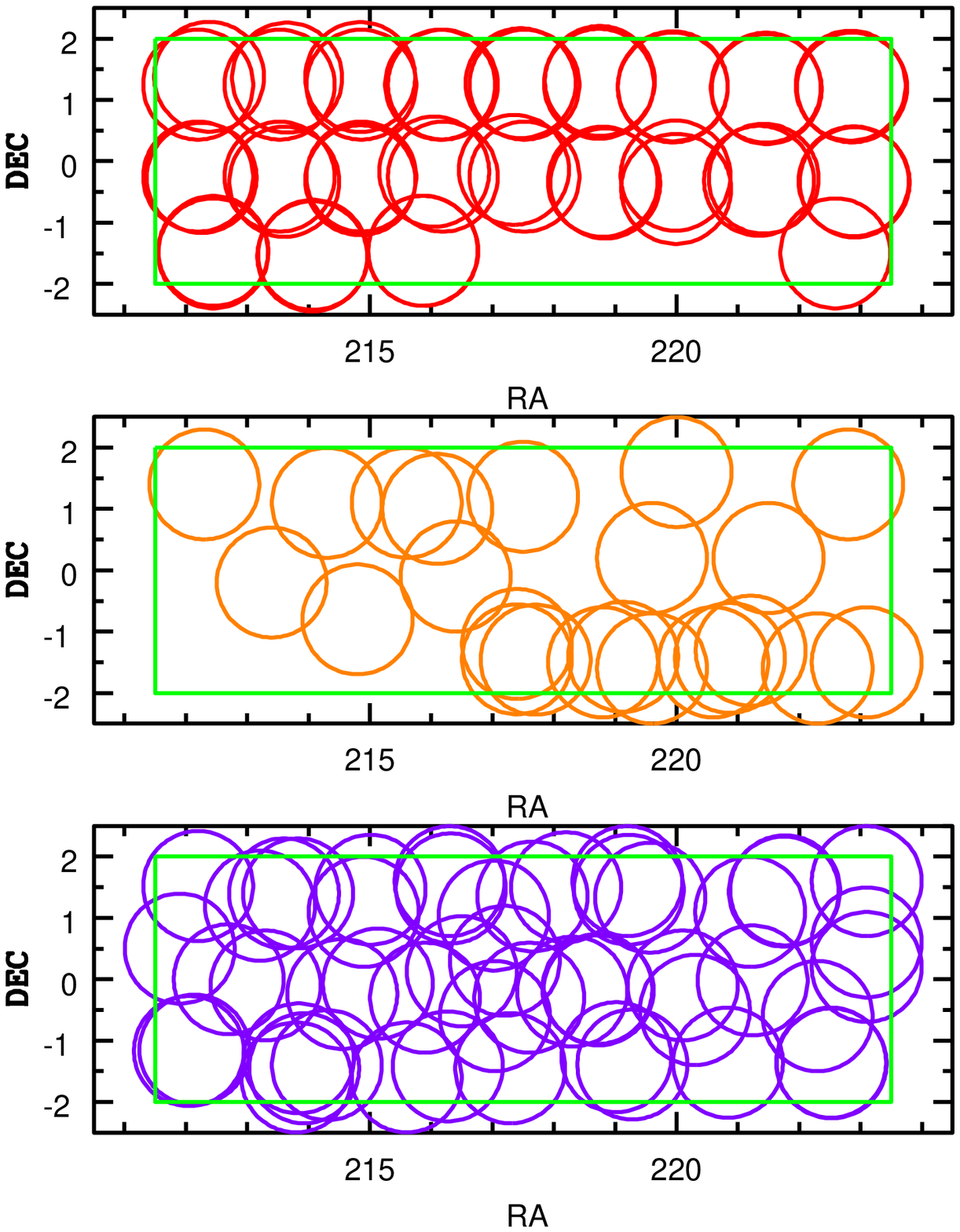,width=7.0cm}}

\caption{Location of the tiles in year 1 (top), year 2 (centre), and
  year 3 (bottom) and for G09 (left), G12 (centre) and G15
  (right). \label{fig:tiles}}
\end{figure*}

\begin{figure*}

\centerline{\psfig{file=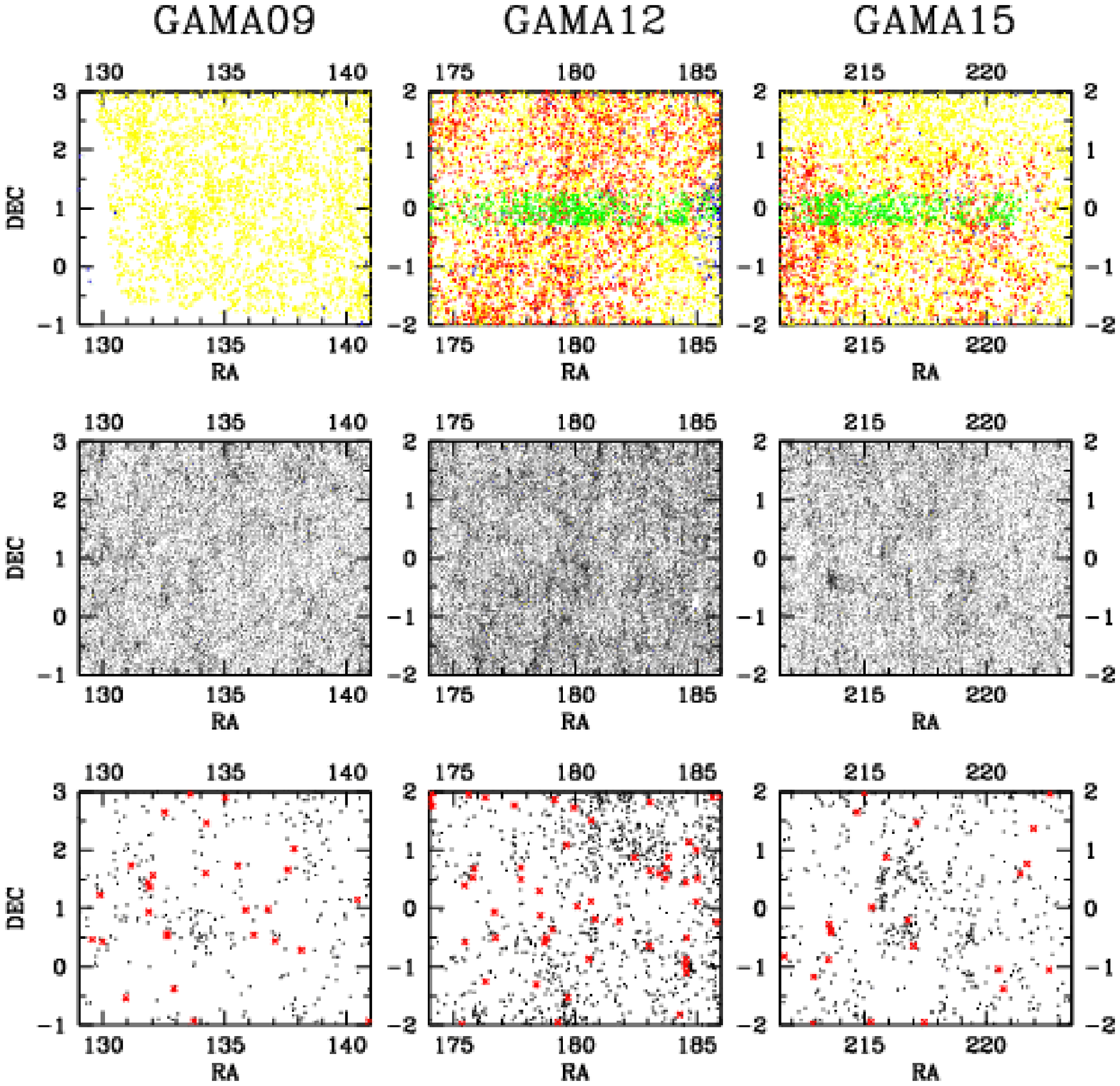,width=\textwidth}}

\vspace{-1.0cm}

\caption{(top) The distribution of pre-existing redshifts (SDSS, yellow;
  2dFGRS, red; MGC, green; other, blue) within the three regions (as
  indicated). (middle) the distributions of redshifts acquired during
  the GAMA Phase I campaign, and (lower) the distribution of targets
  for which redshifts were not obtained (black dots) or were not
  targeted (red crosses). \label{fig:dist}}
\end{figure*}

\begin{figure}
\centerline{\psfig{file=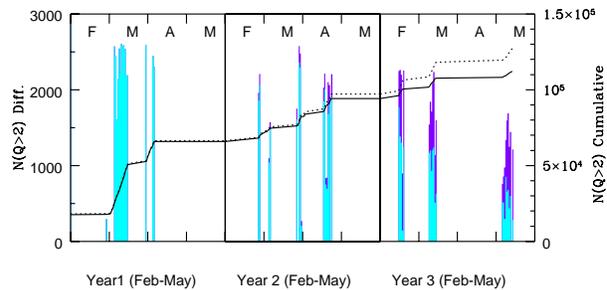,width=\columnwidth}}

\vspace{-4.0cm}

\caption{Progression of the GAMA survey in terms of spectra and
  redshift acquisition for years 1, 2 and 3 as indicated.  The cyan
  line shows the number of spectra obtained per night within our main
  survey limits, mauve shows the number below our flux
  limits (i.e., secondary targets and fillers). The solid line shows
  the cumulative distribution of redshifts within our survey limits
  and the dashed line shows the cumulative distribution of all
  redshifts. The cumulative distributions include pre-existing
  redshifts and are calculated monthly rather than
  nightly. \label{fig:prog}}
\end{figure}

All GAMA 2dF pointings (tiles) were observed during dark or grey time
with exposure times mostly ranging from 3000 to 5000~s (in 3 to 5
exposures) depending on seeing and sky brightness. Observations were
generally conducted at an hour angle of less than 2~hr (the median
zenith distance of the observations is 35$^\circ$) and with the
Atmospheric Dispersion Corrector engaged. We used the 580V and 385R
gratings with central wavelengths of $4800\,$\AA\ and $7250\,$\AA\ in the
blue and red arms, respectively, separated by a
$5700\,$\AA\ dichroic. This set-up yielded a continuous wavelength
coverage of $3720\,$\AA--$8850\,$\AA\ at a resolution of $\approx 3.5\,$\AA\ (in
the blue channel) and $\approx 5.5\,$\AA\ (in the red channel). Every block
of science exposures was accompanied by a flat-field and an arc-lamp
exposure, while a master bias frame was constructed once per observing
run. In each observation fibres were allocated to between 318 and 366
galaxy targets (depending on the number of broken fibres), to at least
20 blank sky positions, to 3--5 SDSS spectroscopic standards, and
to 6 guide stars selected from the SDSS in the range $14.35 < r_{\rm pet} 
< 14.5$ mag (verified via cross-matching with the USNO-B
catalogue to ensure proper motions were below 15 mas/yr). Galaxy
targets were prioritised as indicated in Table~1 of Robotham et
al. (2010). The location of all 392 tile centres are shown in
Fig.~\ref{fig:tiles}.  The change in tiling strategy from a fixed grid
system in year 1 to the greedy tiling strategy in years 2 and 3 is
evident with the latter leading to an extremely spatially uniform
survey (see section~\ref{sec:masks}).  The overall progress of the
survey in terms of objects per night and the cumulative total numbers
are shown in Fig.~\ref{fig:prog} illustrating that typically between
1500 and 2500 redshifts were obtained per night over the three year
campaign.

The data were reduced at the telescope in real time using the
(former\footnote{Note that from 1st July 2010 the Anglo Australian
  Observatory (AAO) has been renamed the Australian Astronomical
  Observatory (AAO), see Watson \& Colless (2010).}) Anglo-Australian
Observatory's 2dfdr software (Croom, Saunders \& Heald 2004) developed
continuously since the advent of 2dF, and recently optimised for
AAOmega. The software is described in a number of AAO
documents\footnote{http://www.aao.gov.au/AAO/2df/aaomega/aaomega\_software.html}. Briefly,
it performs automated tramline detection, sky subtraction, wavelength
calibration, stacking and splicing. Following the standard data
reduction we attempted to improve the sky subtraction by applying a
principal component analysis (PCA) technique similar to that described
by Wild \& Hewett (2005) and described in detail by Sharp \& Parkinson
(2010). We note that $\sim 5$ per cent of all spectra are affected by
fringing (caused by small air gaps between the adhesive that joins a
fibre with its prism), and we are currently studying algorithms that
might remove or at least mitigate this effect.

\subsection{Redshifting}
The fully reduced and PCA-sky-subtracted spectra were initially
redshifted by the observers at the telescope using the code {\sc
  runz}, which was originally developed by Will Sutherland for the
2dFGRS (now maintained by Scott Croom).  {\sc runz} attempts to
determine a spectrum's redshift (i) by cross-correlating it with a
range of templates, including star-forming, E+A and quiescent
galaxies; A, O and M stars; as well as QSO templates; (ii) by fitting
Gaussians to emission lines and searching for multi-line
matches. These estimates are quasi-independent because the strongest
emission lines are clipped from the templates before the
cross-correlation is performed. {\sc runz} then proceeds by presenting
its operator with a plot of the spectrum marking the positions of
common nebular emission and stellar absorption lines at the best
automatic redshift. This redshift is then checked visually by the
operator who, if it is deemed incorrect, may use a number of methods
to try to find the correct one. The process is concluded by the
operator assigning a (subjective) quality ($Q$) to the finally chosen
redshift:

~

\noindent
$Q = 4$: The redshift is certainly correct.

\noindent
$Q = 3$: The redshift is probably correct.

\noindent
$Q = 2$: The redshift may be correct. Must be checked before being
included in scientific analysis.

\noindent
$Q = 1$: No redshift could be found.

\noindent
$Q = 0$: Complete data reduction failure.

~

\noindent
With the above definitions it is understood that by assigning $Q \ge
3$ a redshift is approved as suitable for inclusion in scientific
analysis. Note that $Q$ refers to the (subjective) quality of the {\em
  redshift}, not of the {\em spectrum}.

\subsection{Re-redshifting \label{sec:rered}}
The outcome of the redshifting process described above will not be 100
per cent accurate. It is inevitable that some fraction of the $Q \ge
3$ redshifts will be incorrect. Furthermore, the quality assigned to a
redshift is somewhat subjective and will depend on the experience of
the redshifter. In an effort to weed out mistakes, to quantify the
probability of a redshift being correct, and thereby to homogenise the
quality scale of our redshifts, a significant fraction of our sample
has been independently re-redshifted. This process and its results
will be described in detail by Liske et al. (2010, in prep.). Briefly,
the spectra of all $Q = 2$ and $3$ redshifts, of all $Q = 4$ redshifts
with discrepant photo-$z$'s, and of an additional random $Q = 4$ sample
have been independently re-redshifted, where those conducting the
re-redshifting had no knowledge of the originally assigned redshift or
$Q$. The $Q = 2$ sample was re-examined twice. Overall approximately
one third of the entire GAMA sample was re-evaluated. The results of
the blind re-redshifting process were used to estimate the
probability, for each redshifter, that she/he finds the correct
redshift as a function of $Q$ (for $Q \ge 2$), or that she/he has
correctly assigned $Q = 1$. Given these probabilities, and given the
set of redshift ``opinions'' for a spectrum, we have calculated for
each redshift found for this spectrum the probability, $p_z$, that it
is correct. This allowed us to select the ``best'' redshift in cases
where more than one redshift had been found for a given spectrum. It
also allowed us to construct a ``normalised'' quality scale:

~

\noindent
$nQ$ = 4 if $p_z \ge 0.95$

\noindent
$nQ$ = 3 if $0.9 \le p_z < 0.95$

\noindent
$nQ$ = 2 if $p_z < 0.9$

\noindent
$nQ$ = 1 if it is not possible to measure a redshift from this spectrum.

~

\noindent
Unlike $Q$, whose precise, quantitative meaning depends on the
redshifter who assigned it, the meaning of $nQ$ is homogeneous across
the entire redshift sample.

\begin{figure*}
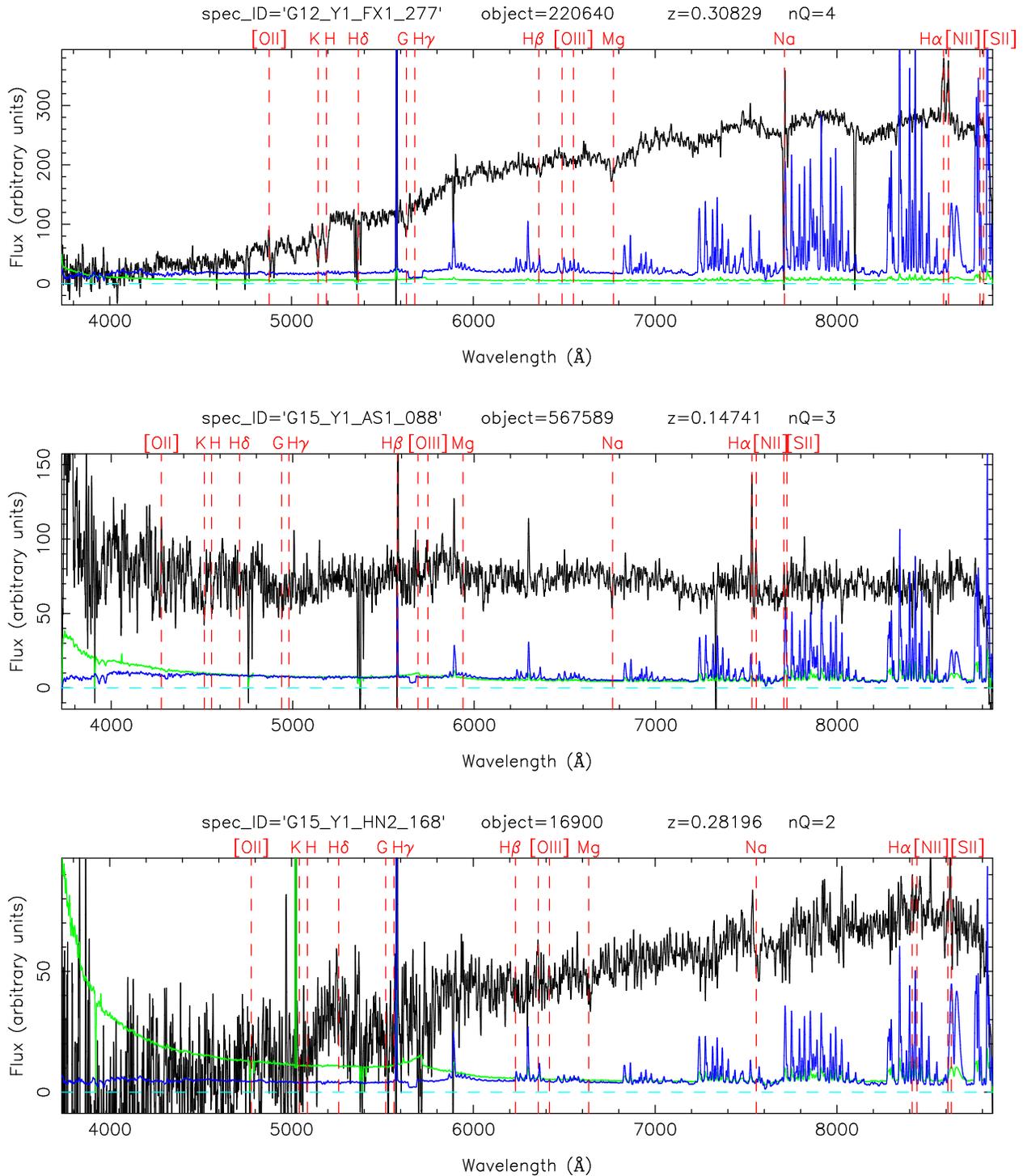


\centerline{\psfig{file=./G12_Y1_FX1_277.ps,angle=-90,width=\textwidth}}

\centerline{\psfig{file=./G15_Y1_AS1_088.ps,angle=-90,width=\textwidth}}

\centerline{\psfig{file=./G15_Y1_HN2_168.ps,angle=-90,width=\textwidth}}

\caption{Examples of spectra with redshift quality $nQ = 4$ (top
  panel), 3 (middle) and 2 (bottom). We show the spectrum (black), the
  1$\sigma$ error (green) and the mean sky spectrum (blue, scaled
  arbitrarily w.r.t. the spectrum). The vertical dashed red lines mark
  the positions of common nebular emission and stellar absorption
  lines at the redshift of the galaxy. The spectra were smoothed with
  a boxcar of width 5 pixels. \label{fig:spectrum}}
\end{figure*}

In Fig.~\ref{fig:spectrum} we show one example each of a spectrum with
$nQ=4$, $nQ=3$ and $nQ=2$.

\subsection{Final redshift sample}
\begin{figure*}
\vspace{-8.0cm}
\centerline{\psfig{file=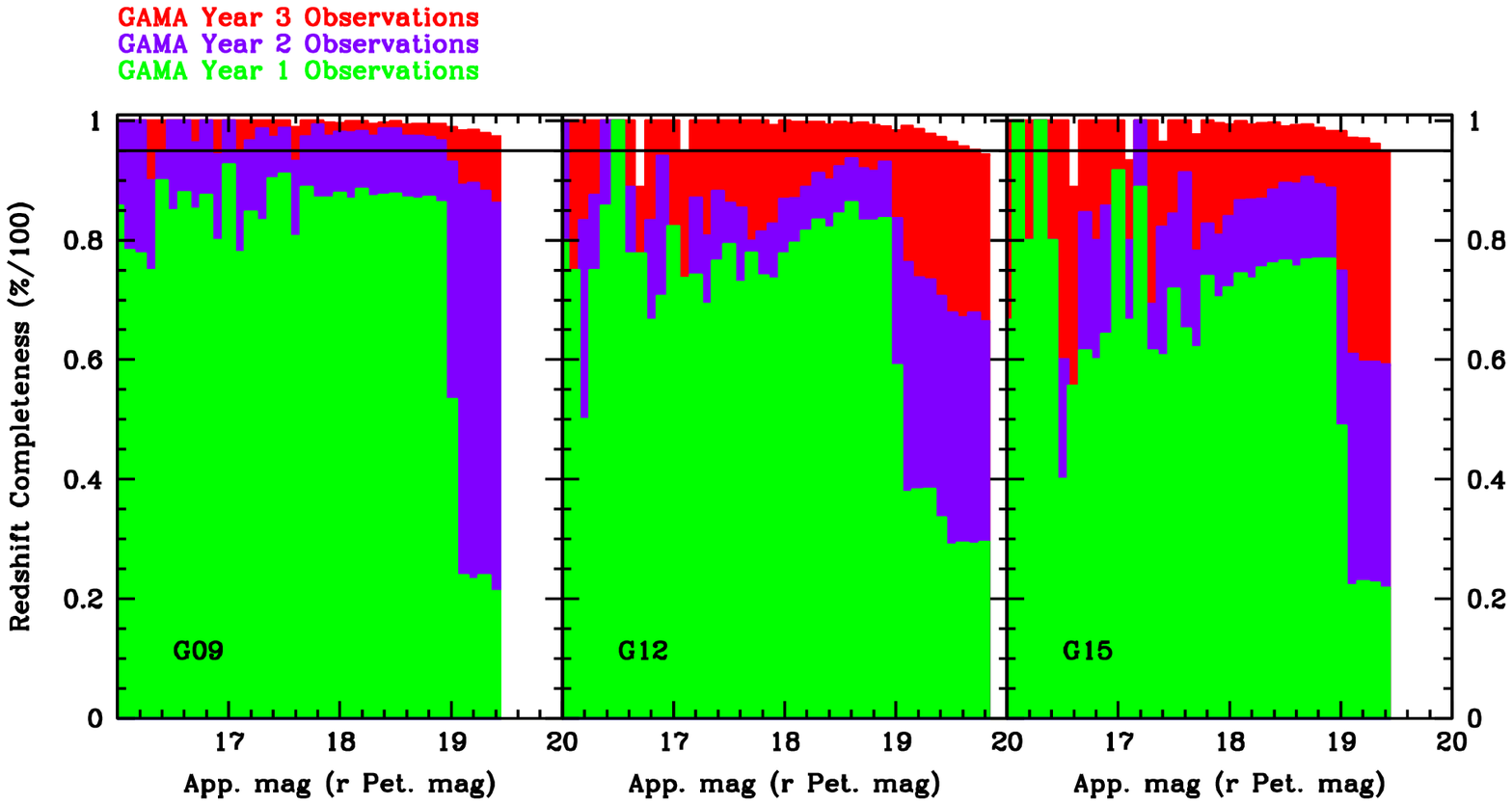,width=\textwidth}}
\caption{Evolution the redshift completeness ($Q \geq 3$) of the GAMA
  survey (main $r$-band selection only) over three years of
  observations showing the progressive build-up towards uniform high
  completeness. The horizontal line denotes a uniform 95 per cent
  completeness. \label{fig:rawcomp}}
\end{figure*}

In the three years of observation completed so far we have observed
392 tiles, resulting in 135\,902 spectra in total (including standard
stars), and 134\,390 spectra of 120\,862 unique galaxy targets. Of
these, 114\,043 have a reliable redshift with $nQ \ge 3$, implying a
mean overall redshift completeness of 94.4 per cent. Restricting the
sample to the $r$-limited Main Survey targets (i.e., ignoring the $K$
and $z$ selection and fainter fillers), we find that the completeness
is $>$ 98 per cent in all three GAMA regions, leaving little room for
any severe spectroscopic bias. Fig.~\ref{fig:rawcomp} shows the
evolution of the survey completeness for the main $r$-band limited
sample versus apparent magnitude across (left-to-right) the three GAMA
regions.

\subsection{Redshift accuracy and reliability}
Quantifying the redshift accuracy and blunder rate is crucial for most
science applications and can be approached in a number of ways. Here
we compare redshifts obtained for systems via repeat observations
within GAMA (Intra-GAMA comparison) and also repeat observations of
objects surveyed by earlier studies (Inter-survey comparison).

\subsubsection{Intra-GAMA comparison}
Our sample includes 974 objects that were observed more than once and
for which we have more than one $nQ \ge 3$ redshift from independent
spectra. The distribution of pairwise velocity differences, $\Delta
v$, of this sample is shown as the shaded histogram in the left panel
of Fig.~\ref{fig:deltaz}. This distribution is clearly not Gaussian
but roughly Lorentzian (blue line), although with a narrower core, and
there are a number of outliers (see below).  Nevertheless, if we clip
this distribution at $\pm$500~km~s$^{-1}$ we find a 68-percentile
range of 185~km~s$^{-1}$, indicating a redshift error $\sigma_v =
65$~km~s$^{-1}$. However, this value is likely to depend on
$nQ$. Indeed, if we restrict the sample to pairs where both redshifts
have $nQ = 4$ (red histogram), we find $\sigma_{v,4} =
60$~km~s$^{-1}$. Our sample of pairs where both redshifts have $nQ=3$
is small (22 pairs) but this yields $\sigma_{v,3} =
101$~km~s$^{-1}$. However, given $\sigma_{v,4}$ we can also use our
larger sample of pairs where one redshift has $nQ=3$ and the other has
$nQ=4$ (green histogram) to obtain an independent estimate of
$\sigma_{v,3} = 97$~km~s$^{-1}$, which is in reasonable agreement.

Defining discrepant redshift pairs as those with $|\Delta v| > $
500~km~s$^{-1}$, and assuming that only one, but not both of the
redshifts of such pairs is wrong, we find that 3.6 per cent of
redshifts with $nQ = 4$ are in fact wrong. This is in reasonable
agreement with the fact that $nQ = 4$ redshifts are defined as those
with $p_z > 0.95$.  However, for $nQ = 3$ we find a blunder rate of
15.1 per cent, which is somewhat higher than expected based on the fact
that $nQ=3$ is defined as $p_z > 0.9$. We surmise that this is likely
to be the result of a selection effect: many of the objects in this sample
are likely to have been re-observed by GAMA because the initial
redshift of the first spectrum was only of a low quality
(i.e.\ $Q=2$). However, subsequent re-redshifting of the initial
spectrum (after the re-observation) may have produced confirmation of
the initial redshift, which will have bumped the redshift quality to
$nQ = 3$. Hence this sample is likely to include many spectra that are
of worse quality, and or spectra that are harder to redshift than the
average $nQ = 3$ spectrum, which will produce a higher blunder rate
for this sample than the average blunder rate for the whole $nQ=3$
sample. Indeed, the median S/N of this sample is 20 per cent lower
than for the full GAMA sample. Note that this may also have an effect
on the redshift accuracies determined above. This will be investigated
in more detail by Liske et al. (2010, in prep).

\subsubsection{Inter-survey comparisons}
\begin{figure*}
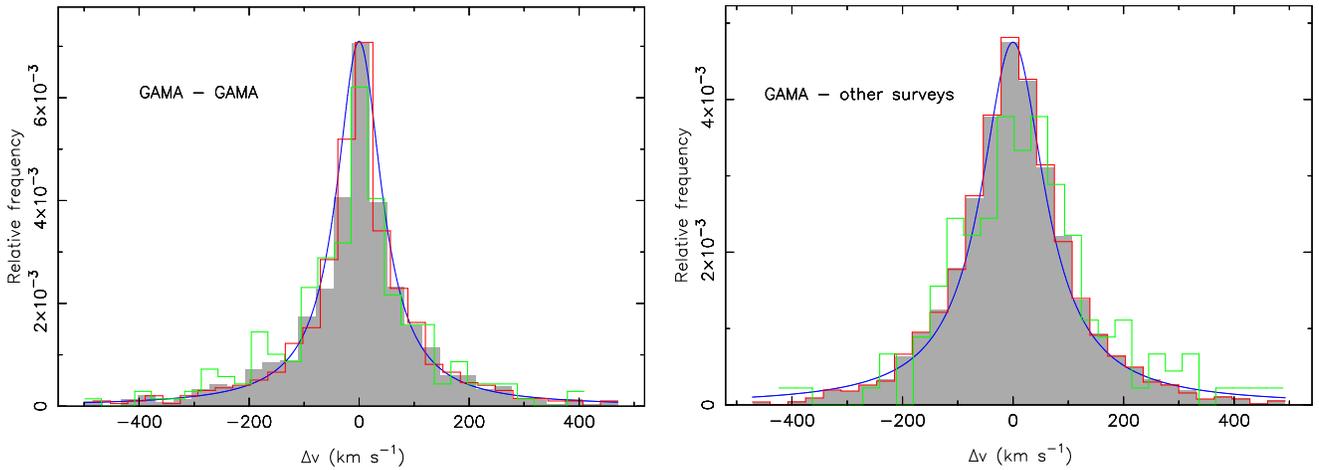


\hspace{-8.75cm}{\psfig{file=./dv_gama.ps,angle=-90,width=8.5cm}}

\vspace{-6.2cm}

\hspace{9.0cm}{\psfig{file=./dv_others.ps,angle=-90,width=8.5cm}}

\caption{Left: The shaded histogram shows the distribution of differences between
the redshifts measured from independent GAMA spectra of the same
objects, where all redshifts have $nQ \ge 3$ (868 pairs from 1718 unique
spectra of 856 unique objects).  The red histogram shows the same for
pairs where both redshifts have $nQ = 4$ (617 pairs from 1216 unique
spectra of 605 unique objects). The green histogram shows the same for
pairs where one redshift has $nQ = 3$ and the other $nQ = 4$ (229 pairs
from 458 unique spectra of 229 unique objects). The blue line shows a
Lorentzian with $\gamma$ = 50~km~s$^{-1}$ for comparison.
Right: The shaded histogram shows the distribution of differences
between the redshifts measured from independent GAMA and non-GAMA
spectra of the same objects, where all redshifts have $nQ \ge 3$ (2533
pairs from 4892 unique spectra of 2359 unique objects). The red
histogram shows the same for pairs where the GAMA redshift has $nQ = 4$
and the non-GAMA redshift $nQ \ge 3$ (2385 pairs from 4618 unique
spectra of 2233 unique objects). The green histogram shows the same for
pairs where the GAMA redshift has $nQ = 3$ (148 pairs from 290 unique
spectra of 142 unique objects). The blue line shows a Lorentzian with
$\gamma$ = 70~km~s$^{-1}$ for comparison. \label{fig:deltaz}}
\end{figure*}

Our sample includes 2522 unique GAMA spectra (with $nQ \ge 3$
redshifts) for objects that had previously been observed by other
surveys (see section~\ref{sec:merge}), for a total of 2671
GAMA--non-GAMA pairs.  The distribution of the velocity differences of
this sample is shown as the shaded histogram in the right panel of
Fig.~\ref{fig:deltaz}.  Approximately 81 per cent of the non-GAMA
spectra in this sample are from the 2dFGRS and the MGC, which were both
obtained with 2dF, using the same set-up and procedures. For the
2dFGRS, Colless et al.\ (2001) quote an average redshift uncertainty of
85~km~s$^{-1}$. Using this value together with the observed
68-percentile ranges of the velocity differences of the
GAMA($nQ=4,3$)--non-GAMA($nQ\ge3$) sample (red and green histograms,
respectively) we find $\sigma_{v,4} = 51$~km~s$^{-1}$ and
$\sigma_{v,3} = 88$~km~s$^{-1}$. These values are somewhat lower than
those derived in the previous section. The most likely explanation for
this is that the inter-survey sample considered here is brighter than
the intra-GAMA sample of the previous section (because of the
spectroscopic limits of the 2dFGRS and MGC). Indeed, the median S/N of
the GAMA spectra in the inter-survey sample is a factor of 1.7 higher
than that of the spectra in the intra-GAMA sample.

As above, we can also attempt to estimate the GAMA blunder rate from
the inter-survey sample. From the GAMA($nQ=4$)--non-GAMA($nQ\ge3$)
sample we find a GAMA $nQ=4$ blunder rate of 5.0 per cent if we assume
that {\em all} redshift discrepancies are due to GAMA
mistakes. However, this is clearly not the case since the blunder rate
improves to 3.0 per cent if we restrict the sample to pairs with $nQ
\ge 4$ non-GAMA redshifts. Similarly, the GAMA $nQ=3$ blunder rate
comes out at 10.1 or 4.2 per cent depending on whether one includes
the pairs with non-GAMA $nQ=3$ redshifts or not, considerably lower
than the corresponding value derived in the previous section.

In summary, it seems likely that neither the intra-GAMA nor the
inter-survey sample are fully representative of the complete GAMA
sample. The former is on average of lower quality than the full sample
while the latter is of higher quality. Hence we must conclude that the
redshift accuracies and blunder rates determined from these samples
are not representative either, but are expected to span the true
values. A full analysis of these issues will be provided by Liske et
al. (2010) following re-redshifting of the recently acquired year 3
data.

\subsection{Merging GAMA with data from earlier redshift surveys}\label{sec:merge}
\begin{figure*}
\centerline{\psfig{file=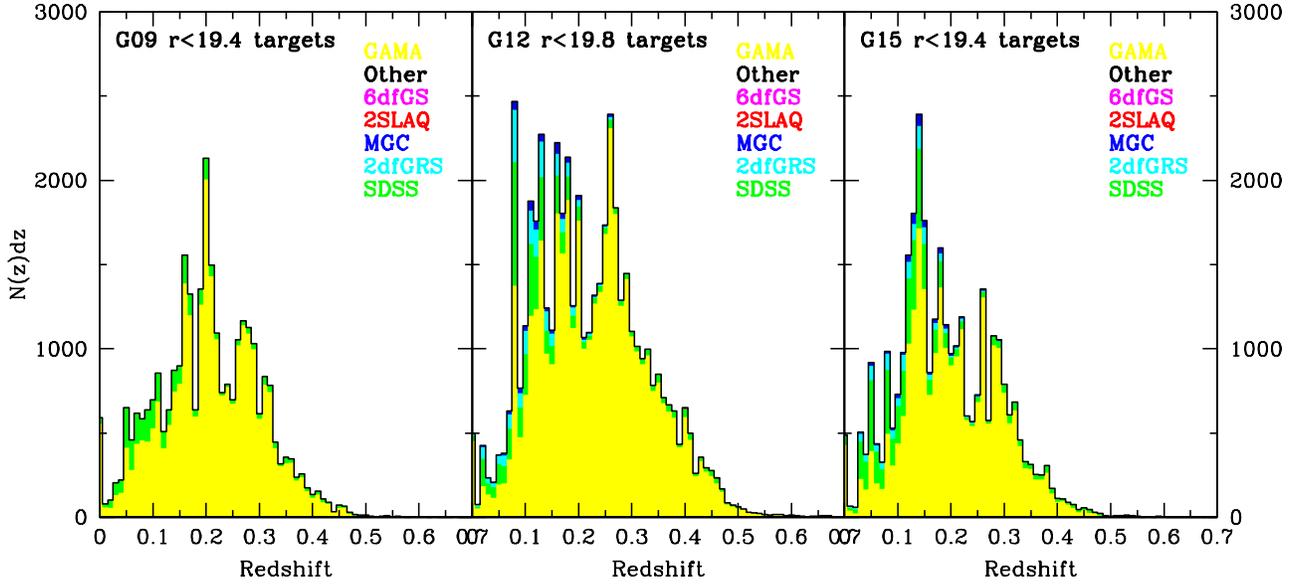,width=\textwidth}}

\vspace{-8.0cm}

\caption{The n(z) distributions of the GAMA regions including new GAMA
  redshifts (shown in yellow) alongside pre-existing redshifts already
  in the public domain as indicated. \label{fig:nz}}
\end{figure*}

The GAMA survey builds upon regions of sky already sampled by a number
of surveys, most notably the SDSS and 2dFGRS but also several others
(as indicated in Table~\ref{tab:otherz}). As the GAMA input catalogue
(Baldry et al.~2010) has taken the pre-existing redshifts into
account, the GAMA data by themselves constitute a highly biased sample
missing 80-90 per cent of bright sources with $r_{\rm pet} < 17.77$
mag and fainter objects previously selected by AGN or LRG surveys.  It
is therefore important for almost any scientific application, outside of
analysing AAOmega performance, to produce extended catalogues that
include both the GAMA and pre-GAMA data. Fig.~\ref{fig:nz} shows the
$n(z)$ distributions of redshifts in the three GAMA blocks to $r_{\rm pet} 
< 19.4$ mag in G09 and G15 and to $r_{\rm pet}<{19.8}$ mag in
G12 colour coded to acknowledge the survey from which they
originate. Table~\ref{tab:otherz} shows the contribution to the
combined redshift catalogue from the various surveys. Note the numbers
shown in Table~\ref{tab:otherz} may disagree with those shown in
Baldry et al.~(2010) as some repeat observations of previous targets
were made.

\begin{table*}
\caption{Contribution from various surveys to the final GAMA database. \label{tab:otherz}}
\begin{center} 
\begin{tabular}{lrrrr} \hline
Survey & G09 & G12 & G15 & Reference or \\
source (Z\_SOURCE) & $r<19.4$ & $r<19.8$ & $r<19.4$ & Acknowledgment \\ \hline
SDSS DR7 & 3190 & 4758 & 5092 & Abazajian et al.~(2009) \\
2dFGRS & 0 & 2107 & 1196 & Colless et al.~(2001)\\
MGC & 0 & 612 & 497 & Driver et al.~(2005) \\
2SLAQ-LRG & 2 & 49 & 13 & Canon et al.~(2006)\\
GAMAz & 26783 & 42210 & 25858 & This paper \\
6dFGS & 7 & 14 & 10 & Jones et al.~(2009) \\
UZC & 3 & 2 & 1 & Falco et al.~(1999)\\ 
2QZ & 0 & 31 & 5 & Croom et al. (2004) \\
2SLAQ-QSO & 1 & 1 & 1 & Croom et al.~(2009)\\
NED & 0 & 2 & 3 & The NASA Extragalactic Database\\
z not known & 347 & 1145 & 591 & --- \\ 
Total & 30333 & 50931 & 33267 & --- \\ \hline
Completeness (\%)& 98.9\% & 97.8\% & 98.0 \% & \\ \hline 
\end{tabular}
\end{center}
\end{table*}

\subsection{Update to visual inspection of the input catalogue}\label{sec:visual}
The GAMA galaxy target catalogue has been constructed in an automated
fashion from the SDSS DR6 (see Baldry et al.~2010) and a number of
manual checks of the data based on flux and size ratios and various
SDSS flags were made. This resulted in 552 potential targets being
expunged from the survey prior to Year~3 (Baldry et
al.~2010). Expunged targets were given \textsc{vis\_class} values of 2
(no evidence of galaxy light) or 3 (not the main part of a galaxy). In
reality, objects with \textsc{vis\_class}=3 were targeted but at a
lower priority (below the main survey but above any filler targets)
with most receiving redshifts by the end of Year~3. About 6 percent of
these were, by retrospective visual inspection and by the difference
in redshift, clearly not part of the galaxy to which they were
assigned.  These 16 objects were added back into the Main Survey
(\textsc{vis\_class}=1).

After the redshifts were all assigned to the survey objects, we
further inspected two distinct categories: very bright objects
($r_{\rm pet} < 17.5$ mag, 9453 objects); and targets for which no
redshift was recovered (2083 objects).  The original visual
classification was made using the SDSS jpg image tools (Lupton et
al.~2004; Neito-Santisteban, Szalay \& Gray 2004).  Here we also
created postage stamp images by combining the $u,r,K$ images, and the
resulting 11536 images were visually inspected by SPD. It became clear
that many of the missed apparently bright galaxies were in fact
probably not galaxies or at least much fainter; and some of the other
apparent targets were also probably much fainter. A \textsc{vis\_class}
value of 4 was introduced meaning ``compromised photometry (selection
mag has serious error)''.  From the inspection, 50 objects were
classified as \textsc{vis\_class}=4 and 40 objects as
\textsc{vis\_class}=2.  Independent confirmation was made by IKB using
the SDSS jpg image tools. In summary, a total of 626 potential targets
identified by the automatic prescription described in Baldry et
al. (2010) were expunged from the survey because of the visual
classification ($2 \le$ \textsc{vis\_class} $\le 4$). These objects
are identified in the {\sf GamaTiling} catalogue available from the
data release website.

\begin{figure}
\centerline{\psfig{file=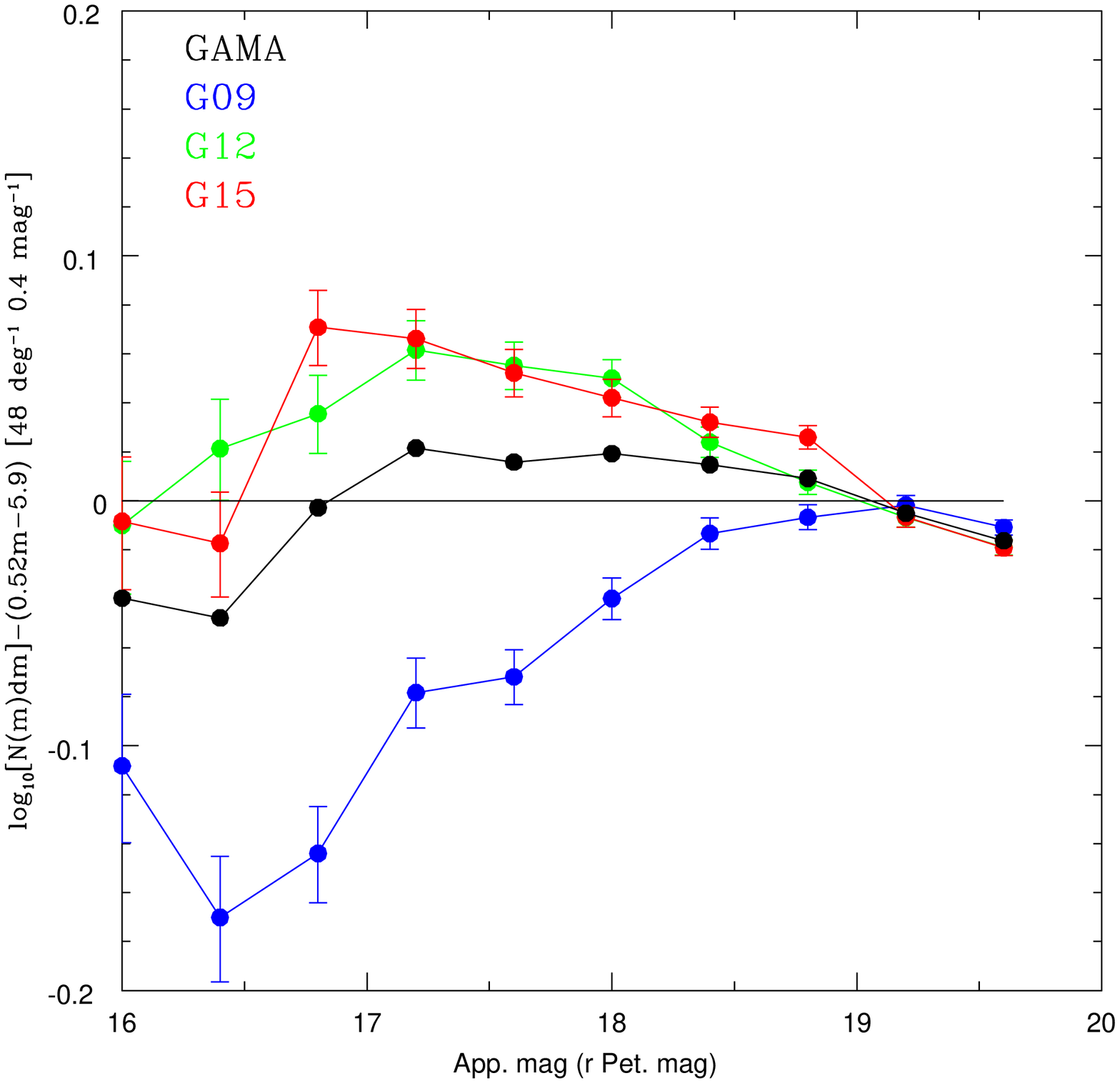,width=\columnwidth}}
\caption{Normalised number-counts as indicated for the three regions
  and the average. Error bars are based on Poisson statistics suggesting
  an additional source of error, most likely cosmic (sample) variance
  between the three GAMA fields. The data becomes consistent only at
  $r_{\rm pet} > 19$ mag. \label{fig:nm}}
\end{figure}

The final input catalogue ({\sf GamaTiling}) therefore constitutes
30331, 50924, and 33261 objects above the Main $r$-band Survey limits
of $r_{\rm pet} < 19.4$, $r_{\rm pet} < 19.8$, and $r_{\rm pet} <
19.4$ mag in G09, G12 and G15 respectively. Fig.~\ref{fig:nm} shows
the normalised galaxy number-counts in these three fields indicating a
significant variation between the three fields and in particular a
significant underdensity in G09 to $r_{\rm pet}=19.0$ mag. This will
be explored further in section~\ref{sec:cv}.

\section{Additional image analysis products}
At this point we have two distinct catalogues: the input catalogue
({\sf GamaTiling}) used by the tiling algorithm, which defines all
viable target galaxies within the GAMA regions; and the combined
redshift catalogue ({\sf GamaRedshifts}), which consists of the
pre-existing redshifts and those acquired by the GAMA Team and
described in detail in the previous section. Our core catalogues
consist of the combination of these two catalogues with three other
catalogues containing additional measurements based on the $ugrizYJHK$
imaging data from SDSS DR6 and UKIDSS LAS archives. These additional
catalogues are: {\sf GamaPhotometry}, {\sf GamaSersic}, and {\sf
  GamaPhotoz} and are briefly described in the sections below (for
full details see Hill et al, 2010a; Kelvin et al. in prep. and
Parkinson et al.~in prep). These five catalogues are then combined and
trimmed via direct name matching to produce the GAMA Core dataset
({\sf GamaCore}), which represents the combined data from these five
catalogues.

\subsection{$ugrizYJHK$ matched aperture photometry ({\sf GamaPhotometry})}\label{sec:swarps}
A key science objective of GAMA is to provide cross-wavelength data
from the far ultraviolet to radio wavelengths. The first step in this
process is to bring together the available optical and near-IR data
from the SDSS and UKIDSS archives. Initially the GAMA team explored
simple table matching, however this highlighted a number of issues,
most notably: differences in deblending outcomes, aperture sizes, and
seeing between the SDSS and UKIDSS datasets. Furthermore as the UKIDSS
$YJHK$ data frames are obtained and processed independently in each
band by The Cambridge Astronomical Survey Unit group there is the
potential for inconsistent deblending outcomes, inconsistent aperture
sizes and seeing offsets between the $YJHK$
bandpasses. Finally both SDSS and UKIDSS LAS measure their empirical
magnitudes (Petrosian and Kron) using circular apertures (see Kron
1980 and Petrosian 1976).  For a partially resolved edge-on system
  this can result in either underestimating flux or adding unnecessary
  sky signal increasing the noise of the flux measurement and
  compounding the deblending issue. To overcome these problems the
  GAMA team elected to re-process all available data to provide
  matched aperture photometry from $u$ to $K$ using elliptical Kron
  and Petrosian apertures defined using the $r$-band data (our primary
  selection waveband). This process and comparisons to the original
  data are outlined in detail in Hill et al.~(2010a) and described
  here in brief as follows:

\begin{figure}
\centerline{\psfig{file=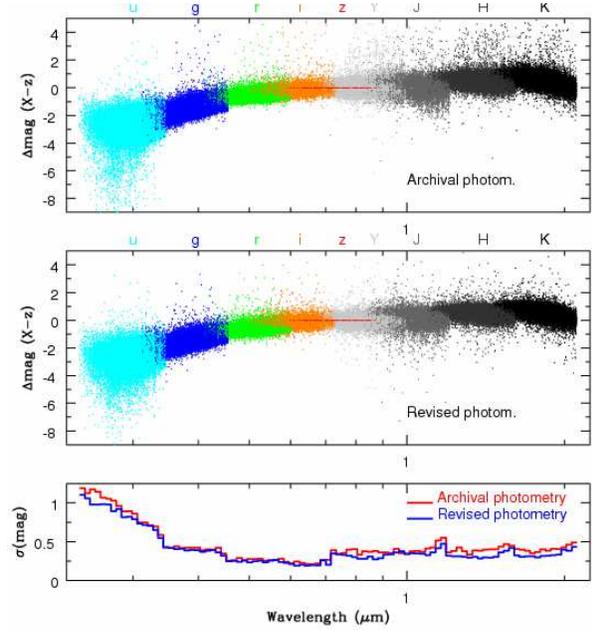,width=\columnwidth}}
\caption{(Upper) all main survey galaxies are plotted according to
  their colour relative to $J$-band at their observed wavelength. The
  data points combine to make a global spectral energy distribute for
  the galaxy population at large. Outlier points are typically caused
  by an erroneous data point in any particular filter. (center) the
  same plot but using the revised photometry. The lower panel shows
  the $5\sigma$-clipped standard deviation for these two distributions
  indicating that in the $uYJHK$ bands the colour distribution is
  quantifiably narrower reflecting an improvement in the photometry.
  providing a cleaner dataset for detailed SED modeling.  Only data
  with photometry in all bands and with secure redshifts less than 0.5
  are shown for clarity.
\label{fig:photom}}
\end{figure}

All available data are downloaded and scaled to a uniform zeropoint in
the strict AB magnitude system (i.e., after first correcting for the
known SDSS to AB offset in $u$ and $z$ bands and converting the UKIDSS
LAS from Vega to AB, see Hill et al.~2010b for conversions). Each
individual data frame is convolved using a Gaussian PSF to yield
consistent 2.0$''$ FWHM PSF measurements for the intermediate flux
stars. The data frames (over 12,000) are then stitched into single
images using the SWARP software developed by the TERAPIX group (Bertin
et al.~2002) resulting in twenty-seven 20~Gb images each of $12
\times 4$ deg$^2$. at $0.4'' \times 0.4''$ pixel sampling. The SWARP
process removes the background using the method described in
SExtractor (Bertin \& Arnouts 1996) using a $256 \times 256$ pixel
mesh. SExtractor is then applied to the $r$-band data to produce a
master catalogue and rerun in each of the other 8 bandpasses in dual
object mode to ensure consistent $r$-defined Kron apertures (2.5$R_K$,
see Kron 1980) from $u$ to $K$. As the data have been convolved to the
same seeing no further correction to the colours are required. For
full details see Hill et al., (2010a). Fig.~\ref{fig:photom} shows the
$u$ to $K$ data based on archival data (upper) and our reanalysis of
the archival data (centre). The data points are plotted as a colour
offset with respect to $z$-band at their rest wavelengths, hence the
data stretch to lower wavelengths within each band with redshift. As
the $k$ and $e$ corrections are relatively small in $z$-band the data
dovetails quite nicely from one filter to the next. The lower panel
shows the standard deviation within each log wavelength interval
indicating the colour range over which the data are spread. In the $U$
and $YJHK$ bands the colour distribution is noticeably narrower for the
revised photometry (blue line) over the archival photometry (red
line). This indicates a quantifiable improvement in the photometry in
these bands over the original archival data with the majority of the
gain occurring for objects in more crowded regions. {\sf
  GamaPhotometry} contains $r$-band detections with matched aperture
photometry for $\sim 1.9$ million objects, $1$ million of which are
matched to SDSS DR6 objects. The photometric pipeline is described in
full in Hill et al. (2010a) and all SWARP images are made publicly
available for downloading at the GAMA website.

\subsection{Sersic profiles ({\sf GamaSersic})}\label{sec:sersic}
By design both Petrosian and Kron magnitude systems recover only a
proportion of a galaxy's flux. For the two most commonly discussed
galaxy profiles, the exponential and the de Vaucouleurs profiles,
traditionally used to describe galaxy discs or spheroids, the Kron
measurement recovers 96 and 90 per cent while the SDSS implementation
of the Petrosian profile recovers 98 per and 83 per cent respectively
(see Graham \& Driver 2005). These two profiles are two specific cases
of the more general S\'ersic profile introduced by S\'ersic (1963,
1968) and more recently reviewed by Graham \& Driver (2005). The
S\'ersic profile (see Eqn.~\ref{eqn:sersic}) is a useful general
description of a galaxy's overall light profile and has also been used
to profile the dark matter distribution in numerical simulations (see
Merritt et al.~2006):
\begin{equation}
I(r)=I_o\exp\left(-(r/\alpha)^{1/n}\right) \label{eqn:sersic}.
\end{equation} 
The S\'ersic model has three primary parameters (see
Eqn.~\ref{eqn:sersic}), which are the central intensity ($I_o$); the
scale-length ($\alpha$); the S\'ersic index ($n$); and two additional
parameters for defining the ellipticity ($\epsilon$) and the position
angle ($\theta$). Note this expression can also be recast in terms of
the effective surface brightness and half-light radius (see Graham \&
Driver 2005 for a full description of the S\'ersic profile). The
S\'ersic index might typically range from $n=0.5$ for diffuse systems
to $n=15$ for concentrated systems. If a galaxy's S\'ersic profile is
known it is possible to integrate this profile to obtain a total
magnitude measurement, and this mechanism was used by the SDSS to
provide model magnitudes by force fitting either an $n=1$
(exponential) or an $n=4$ (de Vaucouleurs) profile to all
objects. Here we attempt to take the next step, which is to derive the
S\'ersic profile with $n$ as a free parameter, in
order to provide total magnitudes for all targets.

As described in full in Kelvin et al.~(2010, in prep.) we use the
GALFIT3 package (Peng et al.~2010) to fit all galaxies to SDSS DR6
$r_{\rm pet} < 22.0$ mag. In brief the process involves the
construction of comparable SWARP'ed images as in
section~\ref{sec:swarps} but without Gaussian PSF convolution (raw
SWARPS). Instead, the PSF at the location of every galaxy is derived
using 20 intermediate brightness stars from around the galaxy as
identified by SExtractor and modeled by the code PSFex (Bertin
priv. comm.). Using the model PSF the 2D S\'ersic profile is then
derived via GALFIT3, which convolves a theoretical profile with the PSF
and minimises the five free parameters. The output S\'ersic total
magnitude is an integral to infinite radius, which is
unrealistic. However relatively little is known as to how the light
profile of galaxies truncate at very faint isophotes with all variants
seen (Pohlen \& Trujillo 2006).

\begin{figure}
\centerline{\psfig{file=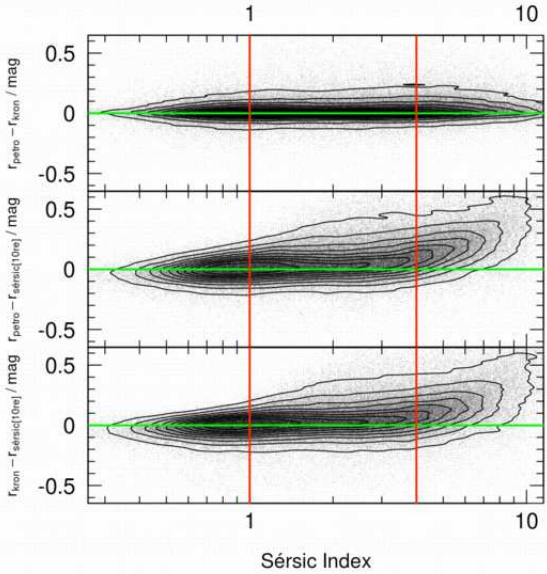,width=\columnwidth}}
\caption{Original SDSS Petrosian magnitudes versus our SExtractor based Kron
magnitudes (top), original SDSS Petrosian magnitudes versus our
GALFIT3 S\'ersic magnitudes (integrated to $10R_e$) (middle), and our
SExtractor Kron versus our GALFIT3 S\'ersic magnitudes (integrated to
$10R_e$) (lower). Contours vary from 10 per cent to 90 per cent in 10
per cent intervals. \label{fig:sersic}}
\end{figure}

In order to calculate an appropriate S\'ersic magnitude it is prudent
to adopt a truncation radius out to which the S\'ersic profile
is integrated. For SDSS model magnitudes exponential ($n=1$) profiles were
smoothly truncated beyond $3R_e$ and smoothly truncated beyond $7R_e$
for de Vaucouleurs ($n=4$) profiles. Here we adopt an abrupt
truncation radius of $10R_e$; for the majority of our data, this
equates to an isophotal detection limit in the $r$-band of $\sim30$
mag/sq arcsec --- the limit to which galaxy profiles have been
explored. We note for almost all plausible values of $n$, the choice
of a truncation radius of $7R_e$ or $10R_e$ introduces a systematic
uncertainty that is comparable or less than the photometric
error. Fig.~\ref{fig:sersic} shows a comparison of the three principle
photometric methods, with $r_{\rm pet}-r_{\rm Kron}$ versus
$\log_{10}(n)$ (upper), $r_{\rm pet}-r_{\rm Sersic10R_e}$ versus
$\log_{10}(n)$ (middle), and $r_{\rm Kron}-r_{\rm Sersic10R_e}$ versus
$\log_{10}(n)$ (lower). Note that it is more logical to compare to
$\log_{10}(n)$ than $n$ as it appears in the exponent of the S\'ersic
profile definition (see review of the S\'ersic profile in Graham \&
Driver 2005 for more information). The top panel indicates that our
Kron photometry (defined in section~\ref{sec:swarps}), reproduces
(with scatter) the original SDSS photometry. As described in the
previous section the motivation for rederiving the photometry is to
provide elliptical matched apertures in $ugrizYJHK$. The middle and
lower plots shows a significant bias between these traditional methods
(Petrosian and Kron) and the new S\'ersic10 $R_e$ magnitude. The
vertical lines indicate the location of the canonical exponential and
de Vaucouleurs profiles. For normal discs and low-$n$ systems the
difference is negligible. However for $n>4$ systems the difference
becomes significant. This is particularly crucial as the high-$n$
systems are typically the most luminous and most massive galaxy
systems (Driver et al.~2006). Hence a bias against this population
can lead to serious underestimation of the integrated properties of
galaxies, e.g., integrated stellar mass or luminosity density (see
Fig.~21 of Hill et al. 2010a, where a 20 per cent increase in the
luminosity density is seen when moving from Petrosian to S\'ersic
magnitudes). The S\'ersic 10$R_e$ magnitudes are therefore doing
precisely what is expected, which is to recover the flux lost via the
Petrosian or Kron photometric systems.

{\sf GamaSersic} contains the five parameter S\'ersic output for all
1.2 million objects with SDSS DR6 $r_{\rm pet}<22$ and full details of
the process are described in Kelvin et al.~(in prep.). At the present
time we only include in our core catalogues the individual profiles
for each object and the r\_SERS\_MAG\_10RE -parameter (see
Table~\ref{tab:gdr1}) while final checks are being conducted. We
release the total correction to allow for an early indication of the
Kron-Total mag offset but caution that these values may
change. Similarly we include for individual objects all profiles with
the full S\'ersic information specified but again caution that these
may change slightly over the next few months as further checks
continue. If colour gradients are small then the $r$-band Kron to
Total magnitude offset given by r\_SERS\_MAG\_10RE can be applied to
all filters.

\subsection{Photometric redshifts ({\sf GamaPhotoz})}\label{sec:photoz}
The GAMA spectroscopic survey is an ideal input dataset for creating a
robust photometric redshift code that can then be applied across the
full SDSS survey region. Unlike most spectroscopic surveys, GAMA
samples the entire SDSS $r$-band selected galaxy population in an
extensive and unbiased fashion to $r_{\rm pet}$ = 19.8 mag -- although
for delivery of robust photo-$z$'s to the full depth of SDSS, it is
necessary to supplement GAMA with deeper data.  Below we briefly
summarize the method, explained in greater detail in Parkinson et
al.~(in prep).

\begin{figure*}

\hspace{-8.0cm}{\psfig{file=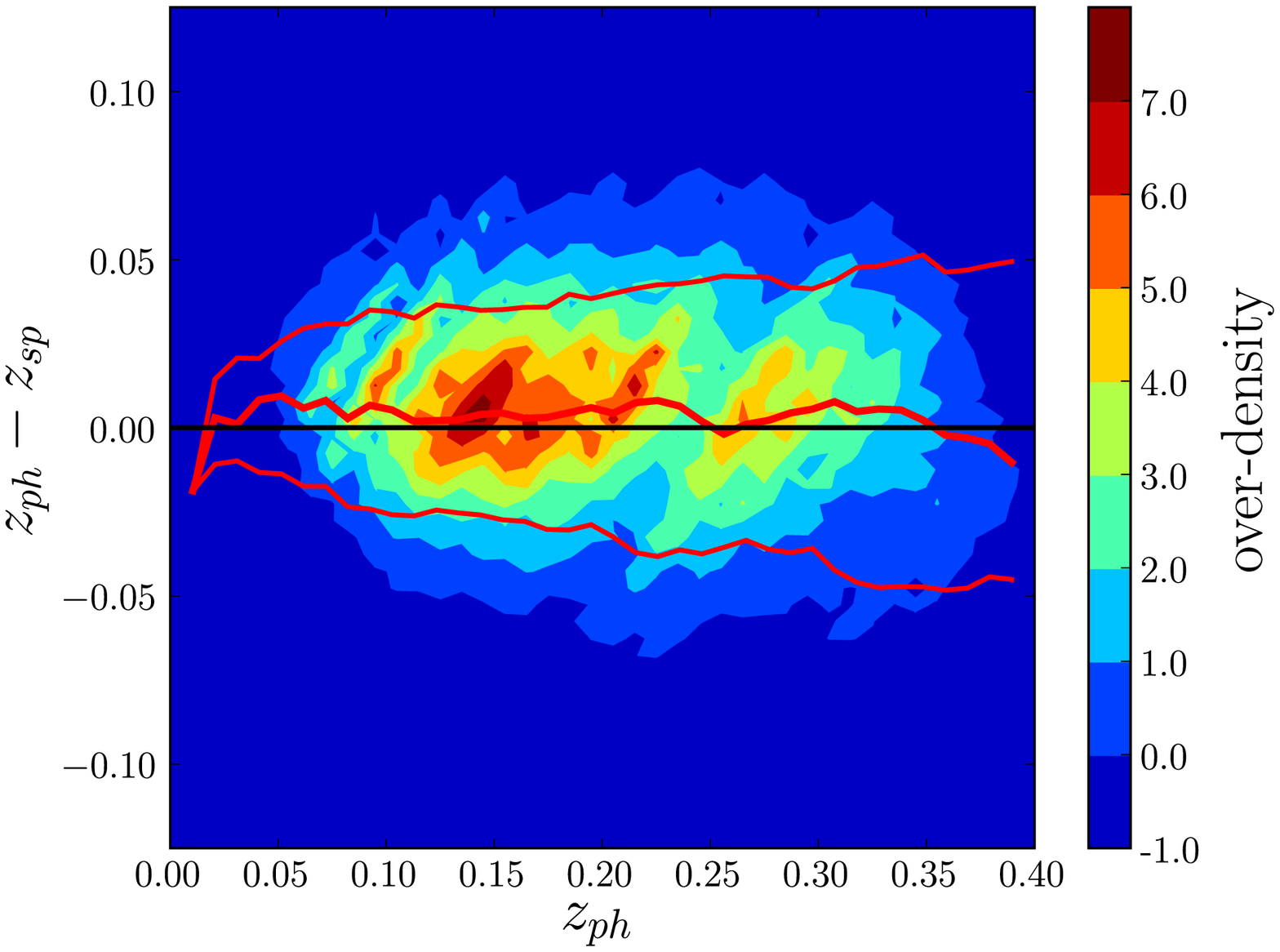,width=8.0cm}}

\vspace{-6.0cm}

\hspace{8.0cm}{\psfig{file=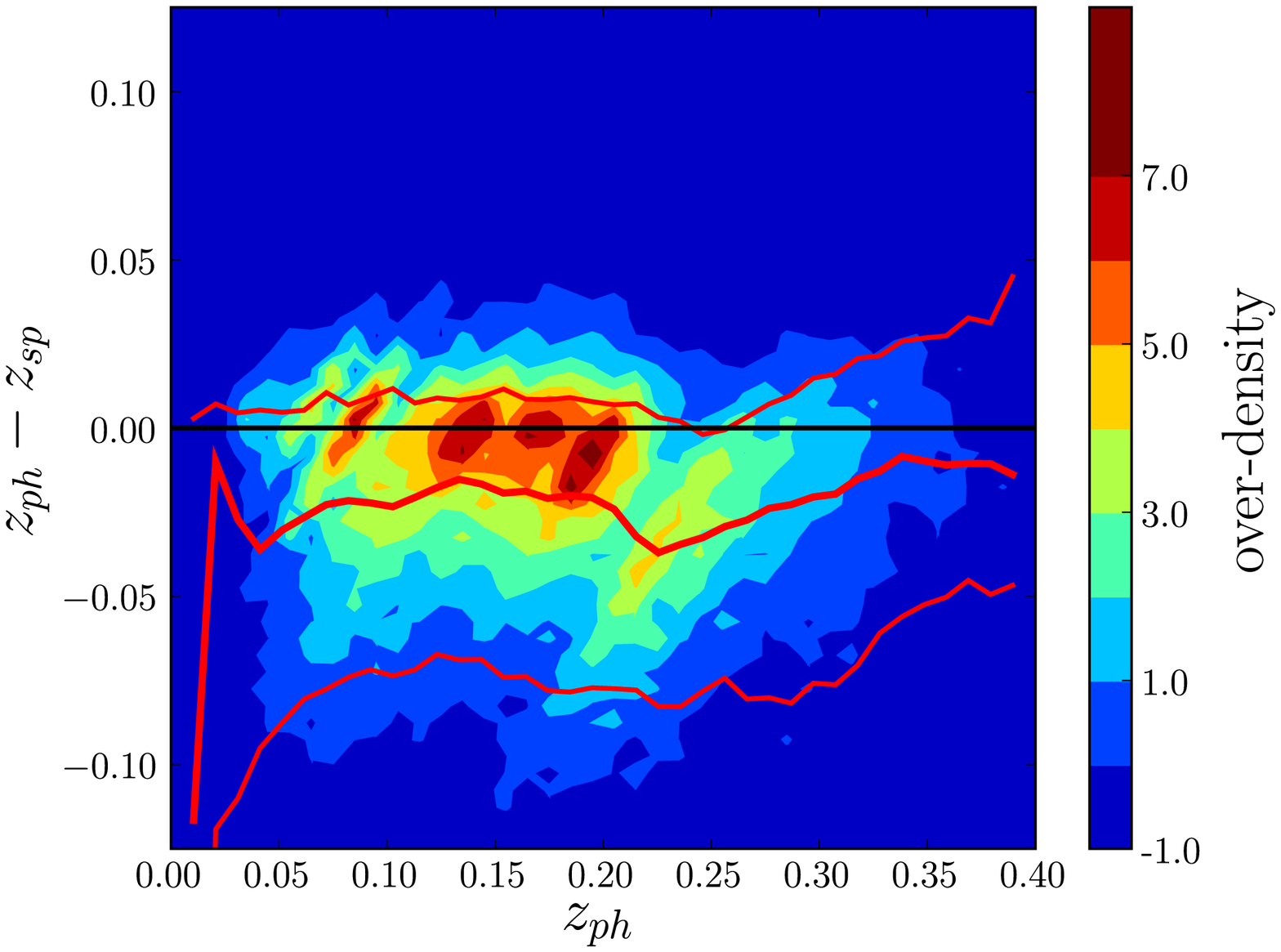,width=8.0cm}}

\caption{(left) GAMA photometric and spectroscopic redshift comparison
  as a function of the photometric redshift for GAMA galaxies, colour
  coded as function of galaxy (over-)density.  The thick red central line
  shows the median, while the outer lines delineate the central 68 per
  cent of the distribution as function of $z_{ph}$.  (right) Same as
  the left panel, but for the SDSS photometric redshift estimate,
  extracted from the Photoz table in SDSS
  DR7.  \label{fig:bias_vs_zph}}
\end{figure*}

We use the Artificial Neural Network code ANNz (Collister \&
Lahav~2004) with a network architecture of N:2N:2N:1, where N=6 is the
number of inputs to the network (5 photometric bands and one radius
together with their errors). In order to develop a photo-$z$ code valid
for the full-SDSS region we use the $ugriz$ \"ubercal calibration of
SDSS DR7 (Padmanabhan al.~2008), together with the best of the De
Vaucouleurs or Exponential half-light radius measurement for each
object (Abazajian et al.~2009). Our training set is composed of all
GAMA spectroscopic redshifts and zCOSMOS spectroscopic redshift (Lilly
et al.~2007), as well as the 30-band COSMOS photometric redshifts
(Ilbert et al.~2009).  With a redshift error of $<0.001$, these are as
good as spectroscopy when it comes to calibrating 5-band photo-$z$'s
from SDSS alone.  The corresponding fractions in the training,
validating and testing sets are 81 per cent, 4 per cent and 15 per
cent respectively, with over 120k galaxies in total. For
COSMOS (and hence zCOSMOS) galaxies, we perform a careful matching to
SDSS DR7 objects.  This large, complete, and representative photo-$z$
training set allows us to use an empirical regression method to
estimate the errors on individual photometric redshifts.  The left
panel of Fig.~\ref{fig:bias_vs_zph} shows that up to $z_{\rm
  ph}\approx0.40$ the bias in the median recovered photo-$z$ is less than
$0.005/(1+z_{\rm ph})$, while beyond it (not shown) the bias is
typically within $0.01/(1 + z_{\rm ph})$. Considering the random
errors on the individual estimates, this systematic bias is
negligible, while still very well quantified. The right panel of
Fig.~\ref{fig:bias_vs_zph} presents the equivalent plot using SDSS DR7
photometric redshifts, as listed in the Photoz table of the SDSS Sky
Server.  In both cases, these plots show only galaxies with $r_{\rm pet}
<19.8$ mag.  This comparison highlights the importance of using
a complete and fully representative sample in the training of
photometric redshifts.

\begin{figure}
\centerline{\psfig{file=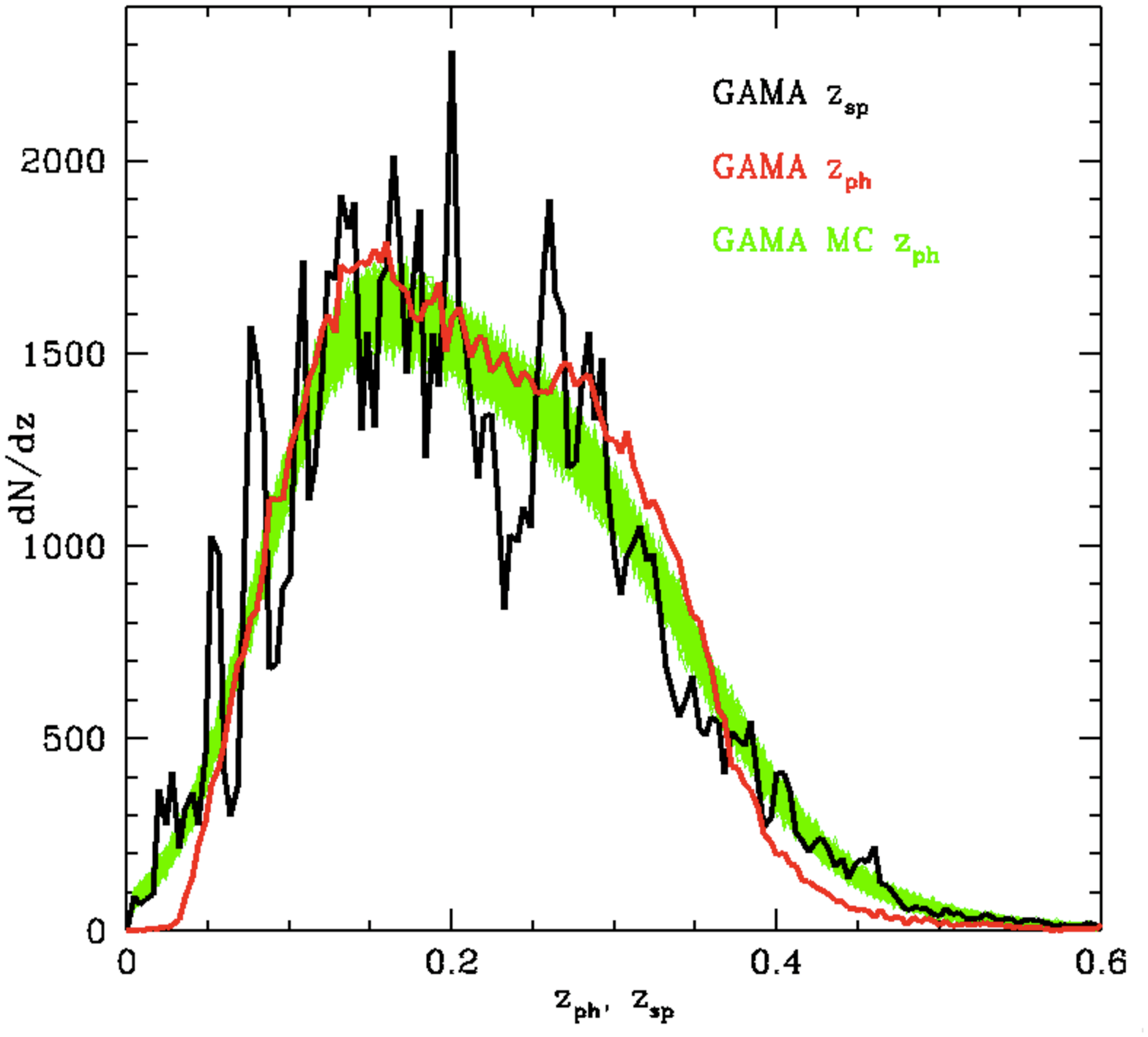,width=\columnwidth}}

\vspace{-2.0cm}

\caption{Spectroscopic redshift distribution of GAMA galaxies
  (black), the corresponding photo-$z$ distribution of GAMA galaxies
  (red) and one thousand Monte-Carlo generated photo-$z$ distributions,
  including the Gaussianly distributed photo-$z$ error
  (green). \label{fig:dndz_photo_z}}
\end{figure}

For an extensive and complete training set such as GAMA, it is
possible to derive a simple and direct method for calculating
photo-$z$ errors. We estimate the photo-$z$ accuracy in a rest-frame
colour magnitude plane with several hundreds of objects per
$0.1\times0.1$ mag bin down to $r_{\rm pet} = 19.8$ mag, and typically
about fifty objects per $0.1\times0.1$ mag bin down to $r_{\rm pet}
\approx 21.5$ mag. This method provides a robust normally distributed
error for each individual object. We note that with this realistic
photo-$z$ error it is possible to correctly recover the underlying
spectroscopic redshift distribution with our ANNz photometric
redshifts as illustrated in Fig.~\ref{fig:dndz_photo_z}.  It is
interesting to note that, without this added error, the raw photo-$z$
histogram is actually narrower than the spectroscopic histogram. This
makes sense, because photo-$z$ calibration yields no bias in true $z$
at given $z_{\rm phot}$: it is thus inevitable that there is a bias in
$z_{\rm phot}$ at given true $z$, which is strong wherever $N(z)$
changes rapidly; this effect can be seen in operation in the ranges
$z<0.05$ and $0.3<z<0.5$.  {\sf GamaPhotoz} contains both photometric
redshifts and errors for all systems to $r_{\rm model} < 21.5$ mag and
the code is currently being applied across the full SDSS region.  We
plan to release our prescription for these improved all-SDSS
photo-$z$'s by the end of 2010. In the meantime, we are open to
collaborative use of the data: contact {\tt gama@gama-survey.org}.

\section{Building and testing the GAMA core catalogues} \label{sec:build}
We can now combine the five catalogues described above to produce the
{\sf GamaCore} catalogue of $\sim1$ million objects, which contains
what we define as our core product to $r<22$
mag. Figs~\ref{fig:coneg09}--\ref{fig:coneg15} uses this catalogue to
illustrate the progression of the redshift survey over the years
through the build-up of the cone plots in the three distinct
regions. As the survey has progressed the fidelity with which
structure can be resolved and groups identified has increased
significantly. In the final maps the survey is over 98\% complete with
a near uniform spatial completeness for all three regions thanks to
the ``greedy'' targeting algorithm (see Robotham et al., 2010). From
{\sf GamaCore} we now extract three science ready subsets:

\begin{figure*}
\centerline{\psfig{file=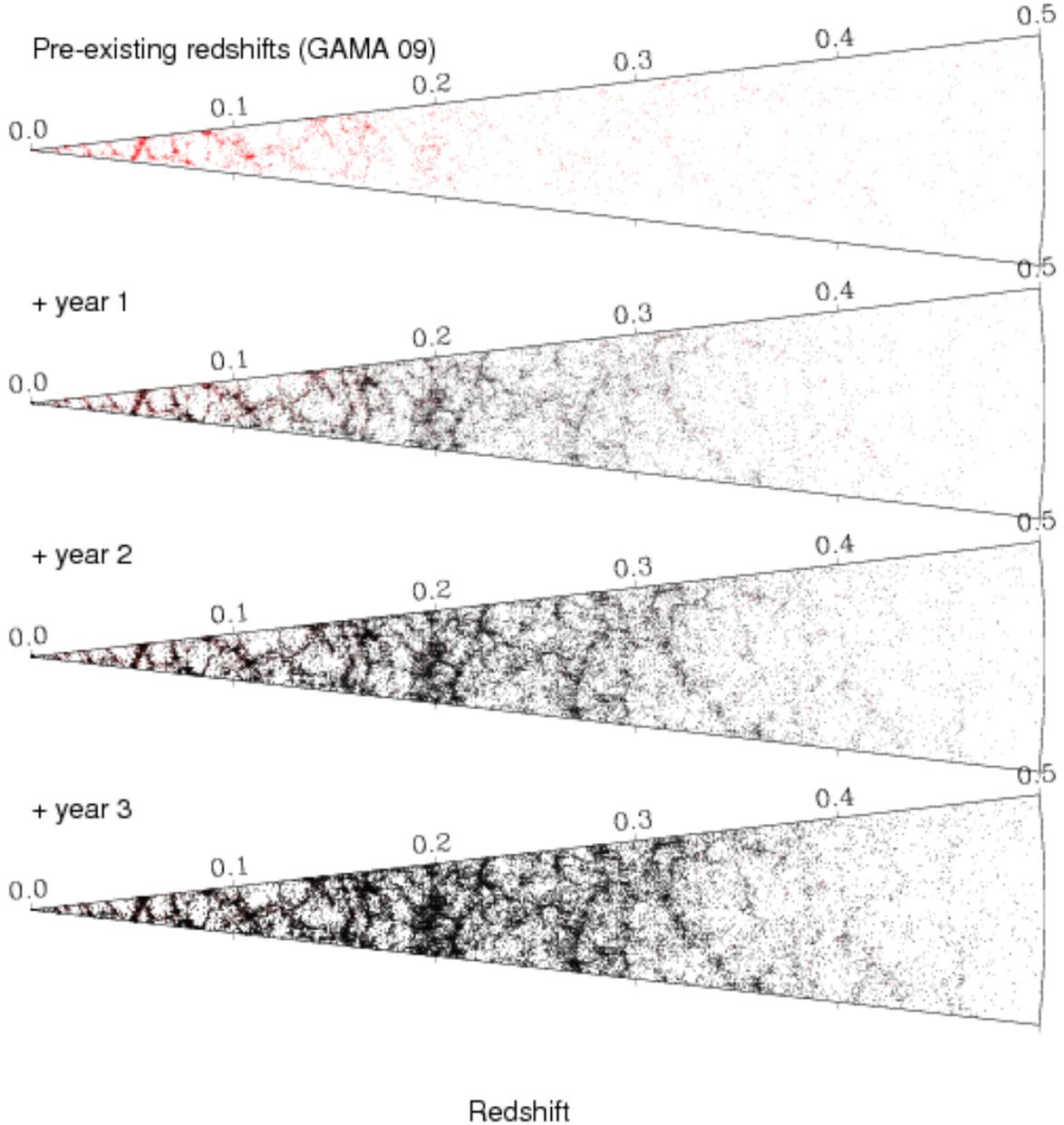,width=\textwidth}}
\caption{Redshift cone diagram for the GAMA 09 region showing (top to
  bottom), pre-existing data, year 1 data release added, year 2 data
  release added, year 3 data release added. \label{fig:coneg09}}
\end{figure*}

\begin{figure*}
\centerline{\psfig{file=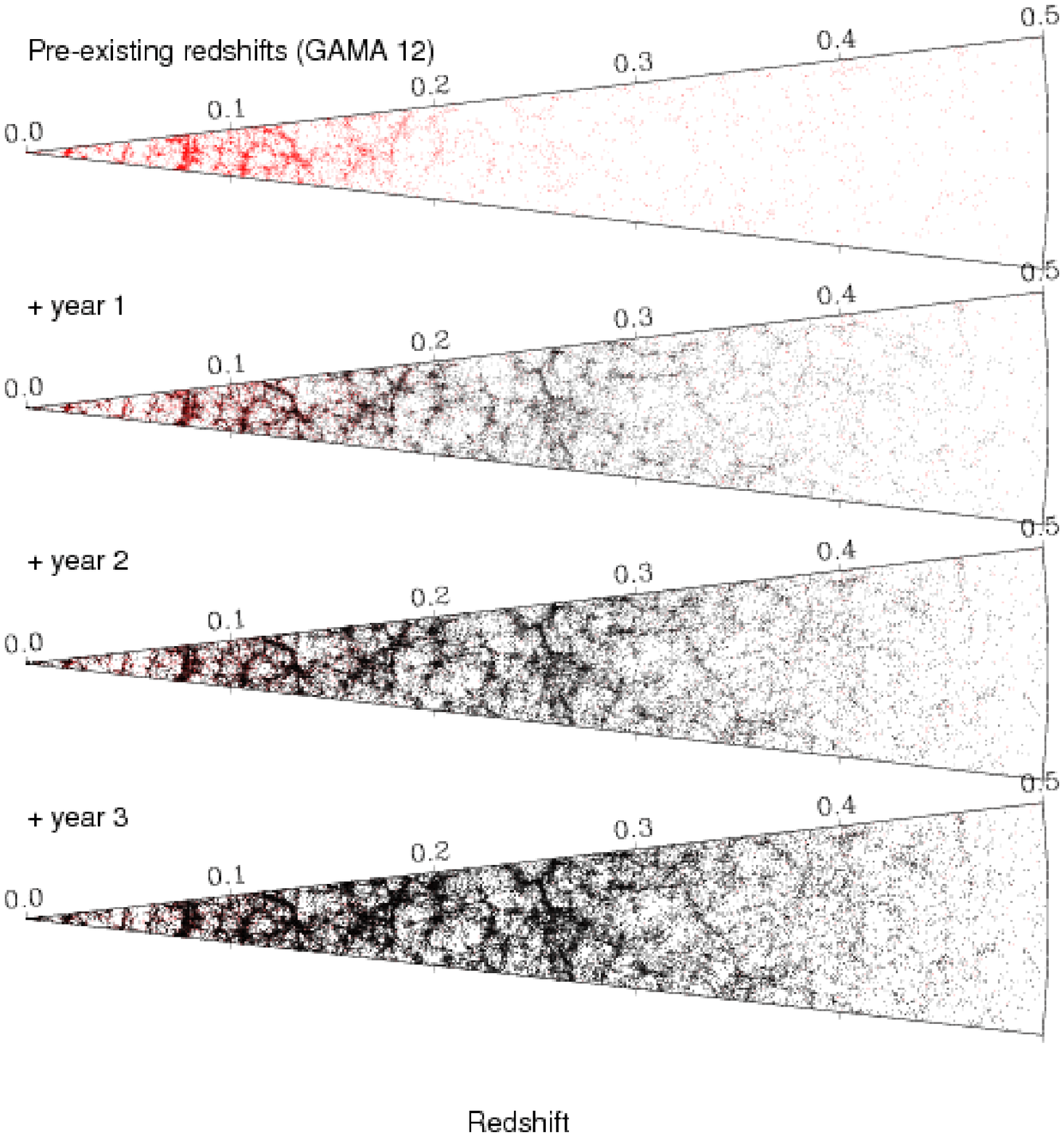,width=\textwidth}}
\caption{Redshift cone diagram for the GAMA 12 region showing (top to
  bottom), pre-existing data, year 1 data release added, year 2 data
  release added, year 3 data release added. \label{fig:coneg12}}
\end{figure*}

\begin{figure*}
\centerline{\psfig{file=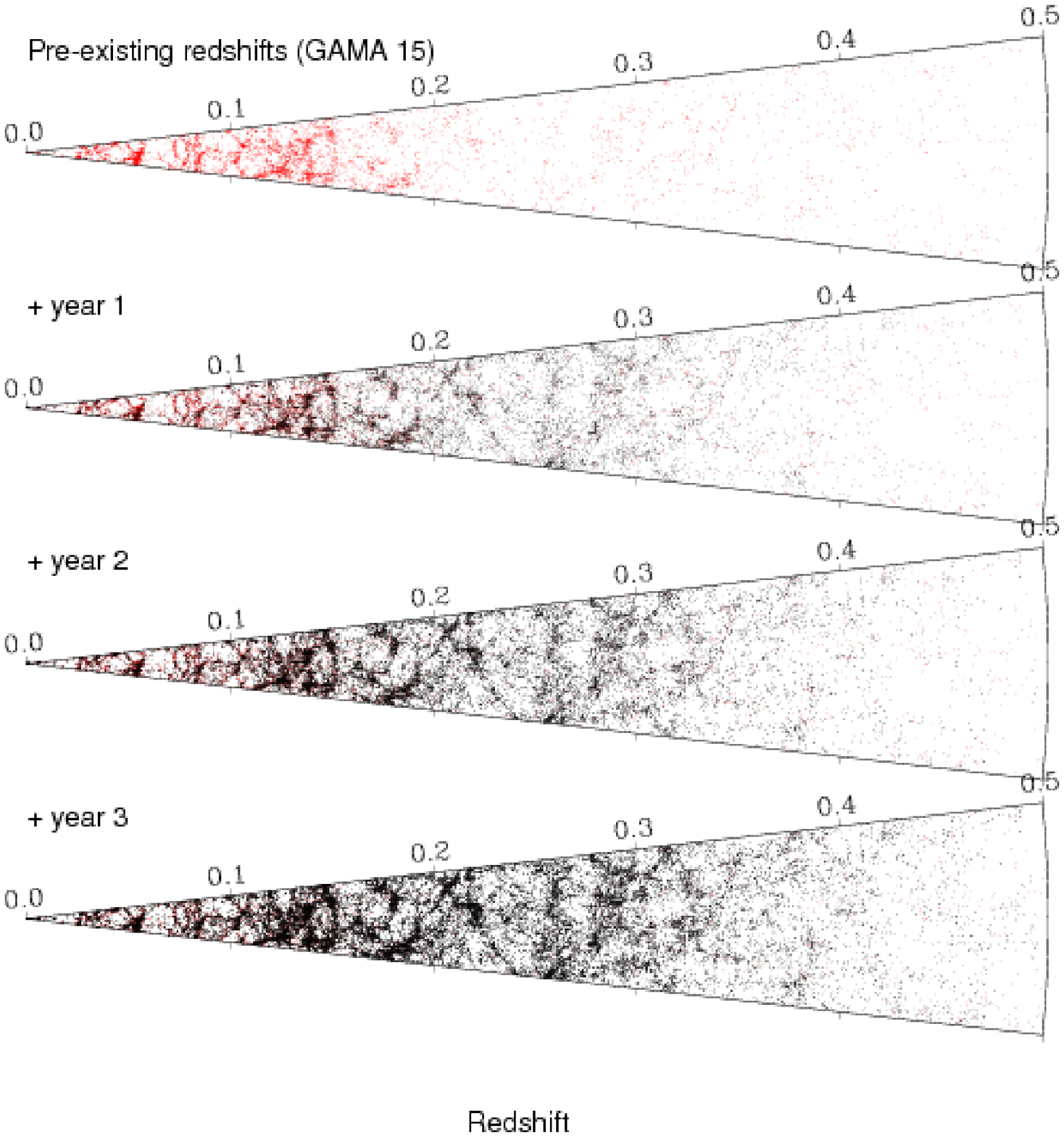,width=\textwidth}}
\caption{Redshift cone diagram for the GAMA 15 region showing (top to
  bottom), pre-existing data, year 1 data release added, year 2 data
  release added, year 3 data release added. \label{fig:coneg15}}
\end{figure*}

(1) GAMA data release 1 ({\sf GamaCoreDR1}), which includes the majority of
spectroscopic data acquired in year 1 with $r_{\rm pet} < 19.0$ mag and a
central strip in G12 ($\delta \pm 0.5^{\circ}$) to $r_{\rm pet} < 19.8$ mag
and which forms our primary public data release product at this time.

(2) GAMA Science ({\sf GamaCoreMainSurvey}), which includes all objects within our
main selection limits and suitable for scientific exploitation by the
GAMA team and collaborators.

(3) GAMA-Atlas ({\sf GamaCoreAtlasSV}), which includes all year 1 and
year 2 data with reliable matches to the H-ATLAS science demonstration
catalogue as described in Smith et al. (2010) and is currently being
used for science exploitation by the H-ATLAS team. For further details
and for the H-ATLAS data release see {\tt http://www.h-atlas.org}

In the following section we explore the
properties of the two primary catalogues ({\sf GamaCoreDR1} and {\sf
  GamaCoreMainSurvey}) and do not discuss {\sf GamaAtlasSV} any
further.

\subsection{Survey completeness}
\begin{figure*}

\vspace{-8.0cm}

\centerline{\psfig{file=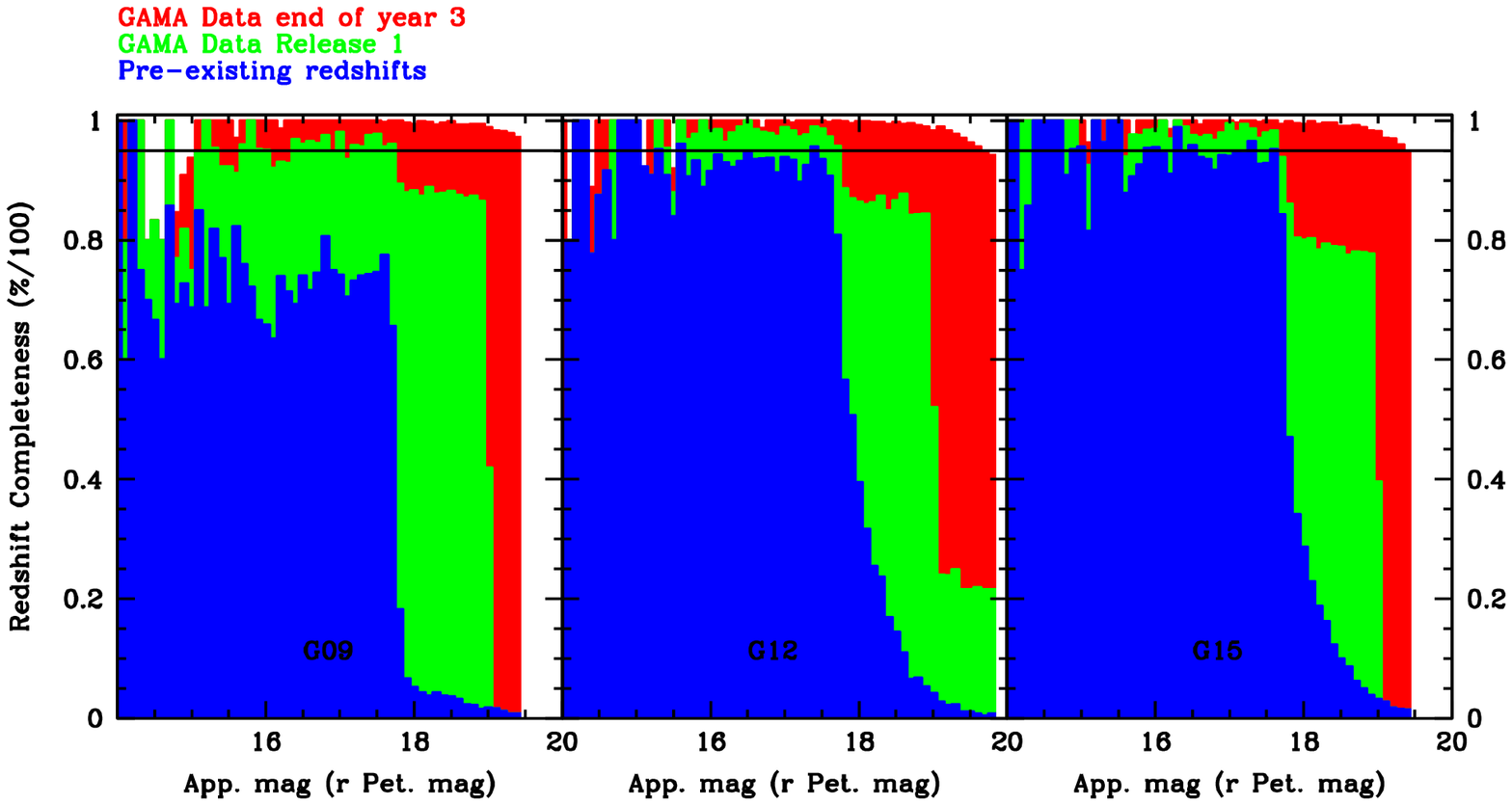,width=\textwidth}}
\caption{The final completeness of the combined redshift catalogues
  prior to survey commencing (blue), after GamaCoreDR1 (green) and
  GamaCoreMainSurvey (red) for G09, G12 and G15 (left to right). In all cases
  the solid black lines shows 95 per cent
  completeness. \label{fig:comp}}
\end{figure*}

In Fig.~\ref{fig:rawcomp} we showed the completeness of the GAMA
redshift survey only. This was given by the number of objects with
secure redshifts divided by those targeted during the three year
spectroscopic campaign. In Fig.~\ref{fig:comp} we show the equivalent
completeness but now for the combined data as described in
section~\ref{sec:merge} for our two principle scientific catalogues
({\sf GamaCoreDR1} and {\sf GamaCoreMainSurvey}). This is given by the
known redshifts in the GAMA regions divided by the known targets in
the GAMA region. The blue regions show the pre-existing redshifts
acquired before GAMA commenced and is mainly dominated by the SDSS DR6
with a spectroscopic limit of $r_{\rm pet} < 17.77$ mag
(cf. Table~\ref{tab:otherz}). {\sf GamaCoreDr1} is shown in green
where the main focus in year 1 was to survey objects brighter than
$r_{\rm pet}=19$ mag with the exception of a central strip in the G12
region (motivated by the desire to assess the completeness function in
preparation for year 2 and 3 observations).

{\sf GamaCoreDR1} is relatively complete to $r_{\rm pet} < 19.0$ mag
and the residual spatial bias can be compensated for by using the
survey masks provided (see section~\ref{sec:masks}). After year 3
observations ({\sf GamaCoreMainSurvey}) one can see 95 per cent or
greater completeness in all apparent magnitude intervals with a smooth
tapering off from ~99 to ~95 per cent completeness in the faintest
0.5mag interval. As all targets without a redshift have now been
visually inspected this fall in completeness is real and obviously needs to be
taken into consideration in any subsequent analysis. However GAMA
remains the most complete survey published to date with the 2dFGRS
achieving a mean survey completeness of 90 per cent to $b_{J}=19.45$
mag, SDSS 90 per cent to $r_{\rm pet}=17.77$ mag, and the MGC 96 per
cent to $B=20$ mag.

\subsection{Survey bias checks}
\begin{figure*}
\centerline{\psfig{file=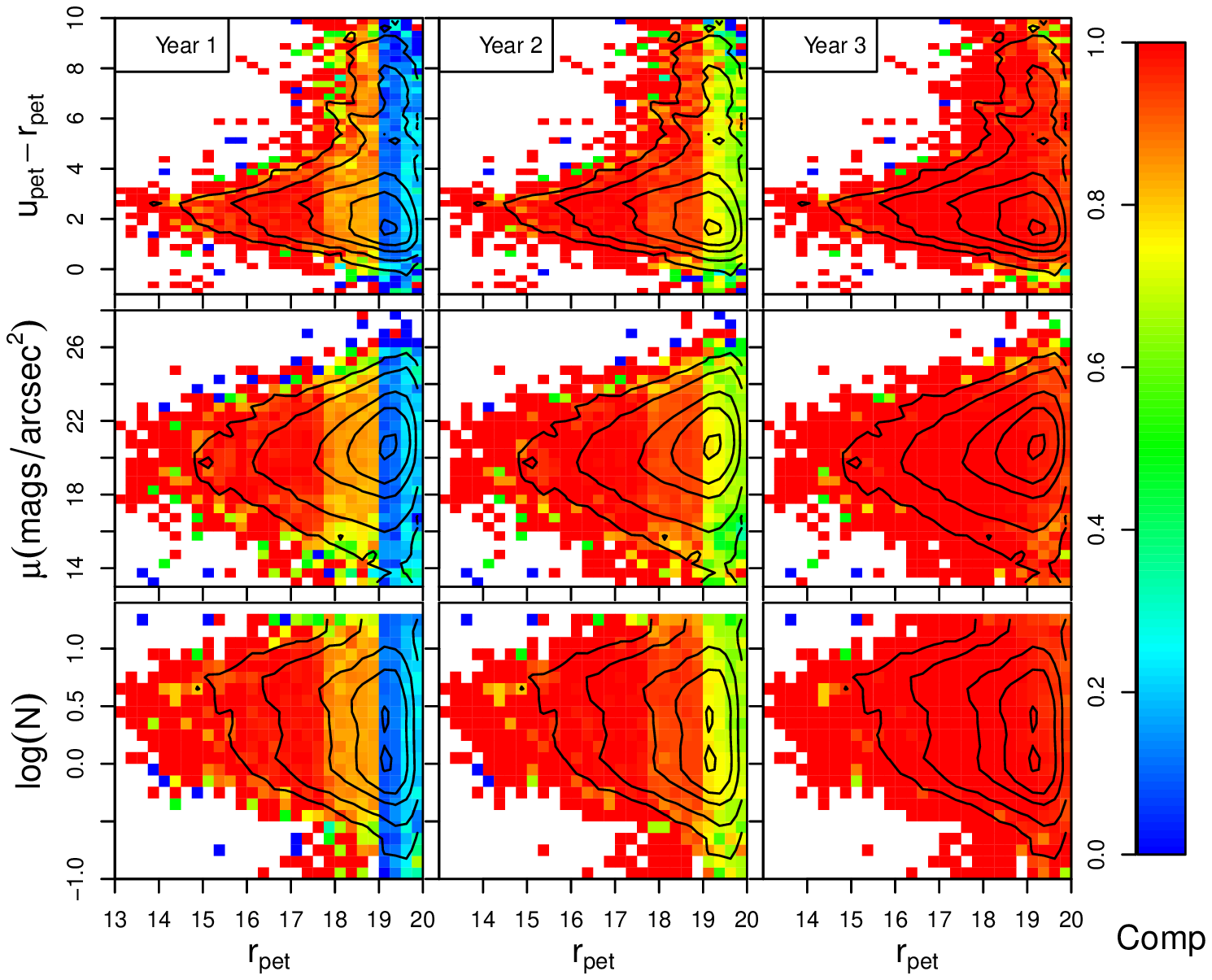,width=\textwidth}}
\caption{Completeness is the bivariate planes of apparent magnitude
  and $u-r$ colour (top), apparent magnitude and effective surface
  brightness (middle), apparent magnitude and S\'ersic index (bottom)
  for GAMA data after year 1 (left), year 2 (centre) and year 3
  (right). Contours show the percentage of galaxies enclosed from out
  to in 99, 95, 75, 50 \& 5 per cent. Apart from the progression in
  apparent magnitude over the three years no other obvious bias is
  evident. \label{fig:bias}}
\end{figure*}

In Fig.~\ref{fig:comp} we see a small but non-negligible completeness
bias with apparent magnitude, hence it is worth checking for any bias
in the obvious parameter space of colour, surface brightness,
concentration, and close pairs. We explore the first three of these in
Fig.~\ref{fig:bias} for year 1 data (left), year 2 data (centre) and
year 3 data (right).  The colour indicates the degree of completeness
in these two-dimensional planes and the contours show the location of
the bulk of the galaxy population. In all three cases there is no
strong horizontal bias that would indicate incompleteness with colour,
surface brightness, or concentration. This is not particularly
surprising given the high overall completeness of the survey, which
leaves little room for bias in the spectroscopic survey, but we
acknowledge the caveat that bias in the input catalogue cannot be
assessed without deeper imaging data. Note that in these plots
surface brightness is defined by: $\mu_{\rm eff}=r_{\rm pet}
+2.5\log_{10}(2 \pi a b)$ where $a$ and $b$ are the major and
minor half-light radii as derived by GALFIT3 (see Kelvin et al.~2010
in prep, for details of the fitting process) and concentration is
defined by $\log_{10}(n)$ where $n$ is the S\'ersic index derived in
section~\ref{sec:sersic}.

\begin{figure}
\centerline{\psfig{file=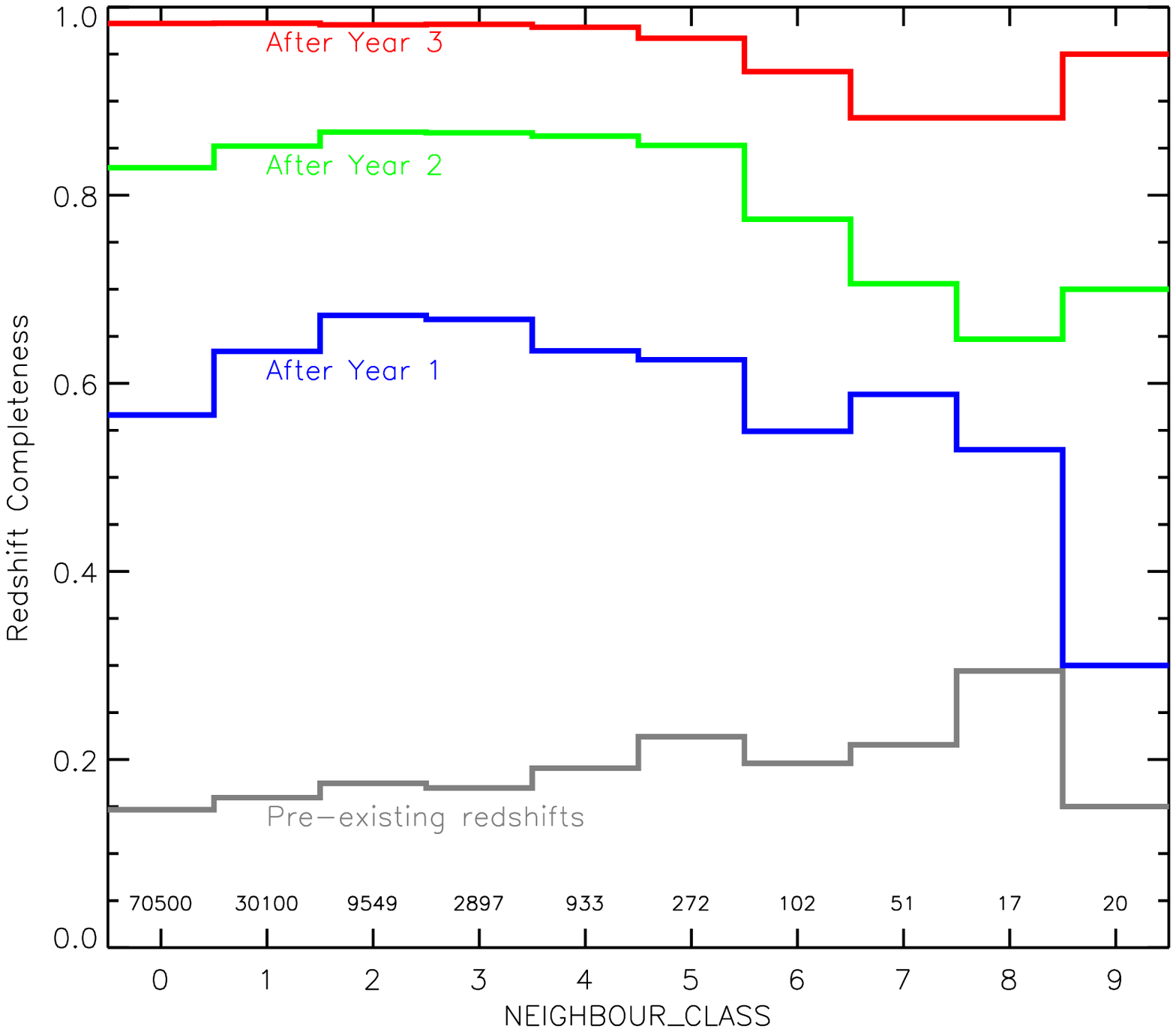,width=\columnwidth}}
\caption{The redshift completeness as a function of Neighbour Class
 (NC). NC is defined for each target as the number of other Main
 Survey targets within $40''$ (numbers with each NC value are annotated
 in the figure). The various lines show the progress in
 resolving clustered objects at the outset and after each year of
 observations for the $r$-limited Main Survey.  Note
 the bias toward NC$>$0 after Year 1 and 2 is the result of increasing
 the priority of clustered targets (Robotham et al. 2010) in order to
 avoid being biased against NC$>$0 after Year 3. The final survey has
 fully resolved almost all close complexes leaving only a minimal
 bias when determining merger rates via close pairs: the redshift
 completeness is $>95$\% for ${\rm NC} \le 5$. \label{fig:pairs}}
\end{figure}

A major science priority of the GAMA programme is to explore close pairs
and galaxy asymmetry (e.g., De Propris et al.~2007). As GAMA is a
multi-pass survey with each region of sky being included in 4-6 tiles
(see. Fig.~\ref{fig:tiles}), the close pair biases due to minimum
fibre separations, which plague the 2dFGRS, SDSS, and MGC studies, are
overcome.  Fig.~\ref{fig:pairs} shows the redshift completeness versus
neighbour class. The neighbour class of any galaxy is defined (see
Baldry et al.~2010) as the number of target galaxies within a $40''$
radius.  The nominal 2dF fibre-collision radius. The figures shows the
progress towards resolving these complexes at the outset and after
each year the survey has been operating. All low neighbour class
objects are complete with only a few redshifts outstanding in a small
fraction of the higher neighbour class complexes. The data
therefore represents an excellent starting point for determining
merger rates via close pair analyses.

\subsection{Overdensity/underdensity of the GAMA regions}\label{sec:cv}
\begin{figure}
\centerline{\psfig{file=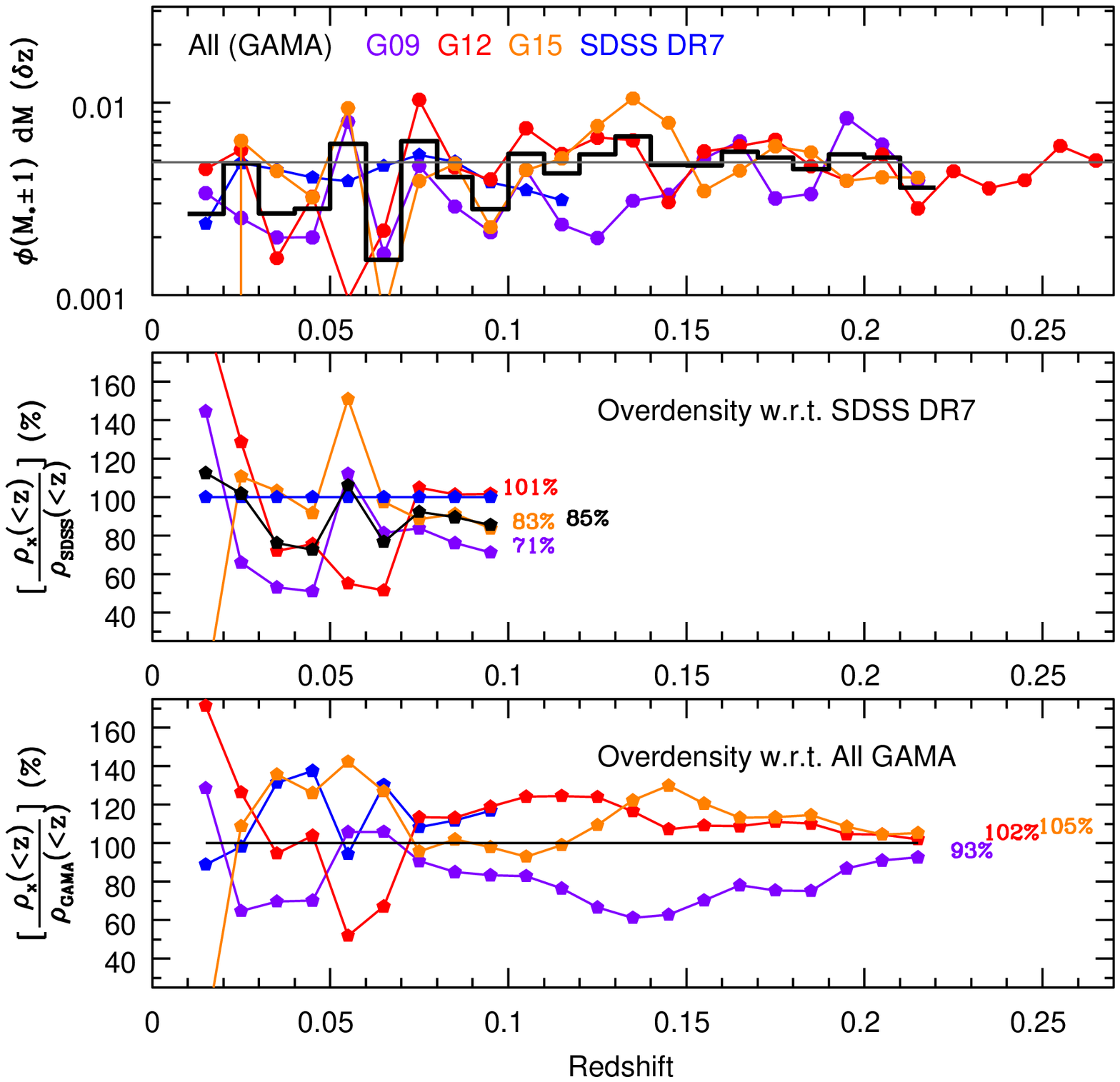,width=\columnwidth}}
\caption{(upper) The differential number-density of $M_*\pm1.0$ mag
  galaxies in redshift intervals of $0.01$.  (middle) the overdensity of
  $M_*\pm1.0$ mag galaxies out to the specified redshift for the
  survey indicated by line colour relative to that seen in a 5000
  sq.deg. region of SDSS DR7. (lower) the overdensity of $M_*\pm1.0$
  mag galaxies out to the specified redshift for the survey indicated
  relative to that seen over all three GAMA regions.
\label{fig:cv}}
\end{figure}

All galaxy surveys inevitably suffer from cosmic variance, more
correctly stated as sample variance. Even the final SDSS, the largest
redshift survey to date, suffers an estimated residual 5 per cent
cosmic variance to $z<0.1$ (see Driver \& Robotham 2010) --- mainly
attributable to the ``Great Wall'' (see Baugh et al.~2004; Nichol et
al.~2006). As the GAMA regions lie fully within the SDSS it is
possible to determine whether the GAMA regions are overdense or
underdense as compared to the full SDSS over the unbiased redshift
range in common ($z<0.1$) and relative to each other at higher
redshifts where the SDSS density drops due to incompleteness imposed
by the $r_{\rm pet} < 17.77$ mag spectroscopic
limit. Fig.~\ref{fig:cv} shows results with the upper panel showing
the density of $M^*\pm1.0$ galaxies (defined as $M^*-5\log h=-20.81$
mag) in $0.01$ redshift intervals, for a 5000 sq.deg. region of the
SDSS DR7 (blue), and for the three GAMA blocks (as indicated) and the
sum of the three regions (solid black histogram).  The $k$ and $e$
corrections adopted and the methodology, including the SDSS sample
construction, are described in Driver \& Robotham (2010) but are not
particularly important as we are not exploring trends with z but
rather the variance at fixed z. The error bars shown are purely
Poisson and so discrepancies larger than the error bars are indicative
of cosmic variance induced by significant clustering along the
line-of-sight. The middle and lower panels show the sum of these
density fluctuations relative to either SDSS (middle) or the average
of the three GAMA fields (lower). Note that these panels (middle,
lower) now show the density fluctuation {\it out to} the specified
redshift as opposed to {\it at} a specific redshift (top), i.e., the
cosmic variance. We can see that out to $z=0.1$ the three GAMA fields
are overall 15 per cent under dense with respect to a 5000
sq.deg. region of SDSS DR7 (NB: the volume surveyed by the SDSS
comparison region is $\sim 1.3 \times 10^{7} h^{-3}$Mpc$^{3}$). This
is extremely close to the 15 per cent predicted from Table~2 of Driver
\& Robotham (2010). Beyond $z=0.1$ one can only compare internally
between the three GAMA fields. The inference from Fig.~\ref{fig:cv} is
that for any study at $z<0.2$ the cosmic variance between the three
regions is significant with G09 in particular being under-dense for
all redshifts $z<0.2$ when compared to the other two regions (NB: the
volume surveyed within the combined GAMA regions to $z<0.2$ is $\sim 2.8 \times
10^6 h^{-3}$Mpc$^{3}$). It is therefore important when considering any
density measurement from GAMA data to include the cosmic variance
errors indicated in Fig.~\ref{fig:cv}. For example a luminosity
function measured from the G09 region only out to $z=0.1$ would need
to be scaled up by $\times 1.41$ etc.

\section{Masks}\label{sec:masks}
For accurate statistical analysis of GAMA it is essential to have a full
understanding of the criteria that define its parent photometric catalogue,
and also of the spatial and magnitude-dependent completeness of the
redshift catalogue. For this purpose we have defined three functions
characterizing this information as a function of position on the sky
and magnitude selection. Here we present briefly the two most
important ones, i.e. the survey imaging mask function and the redshift
completeness mask relative to the main $r_{\rm pet}$ survey
limits. The combination of both functions is key for any spatially
dependent measurements based on GAMA data, and in particular on GAMA
year 1 data. Additional survey completeness masks for other selections
(e.g., $z$, $K$) or any combination of are available to the Team and
will be released shortly.

\begin{figure*}

\vspace{-2.0cm}

\hspace{-8.0cm}{\psfig{file=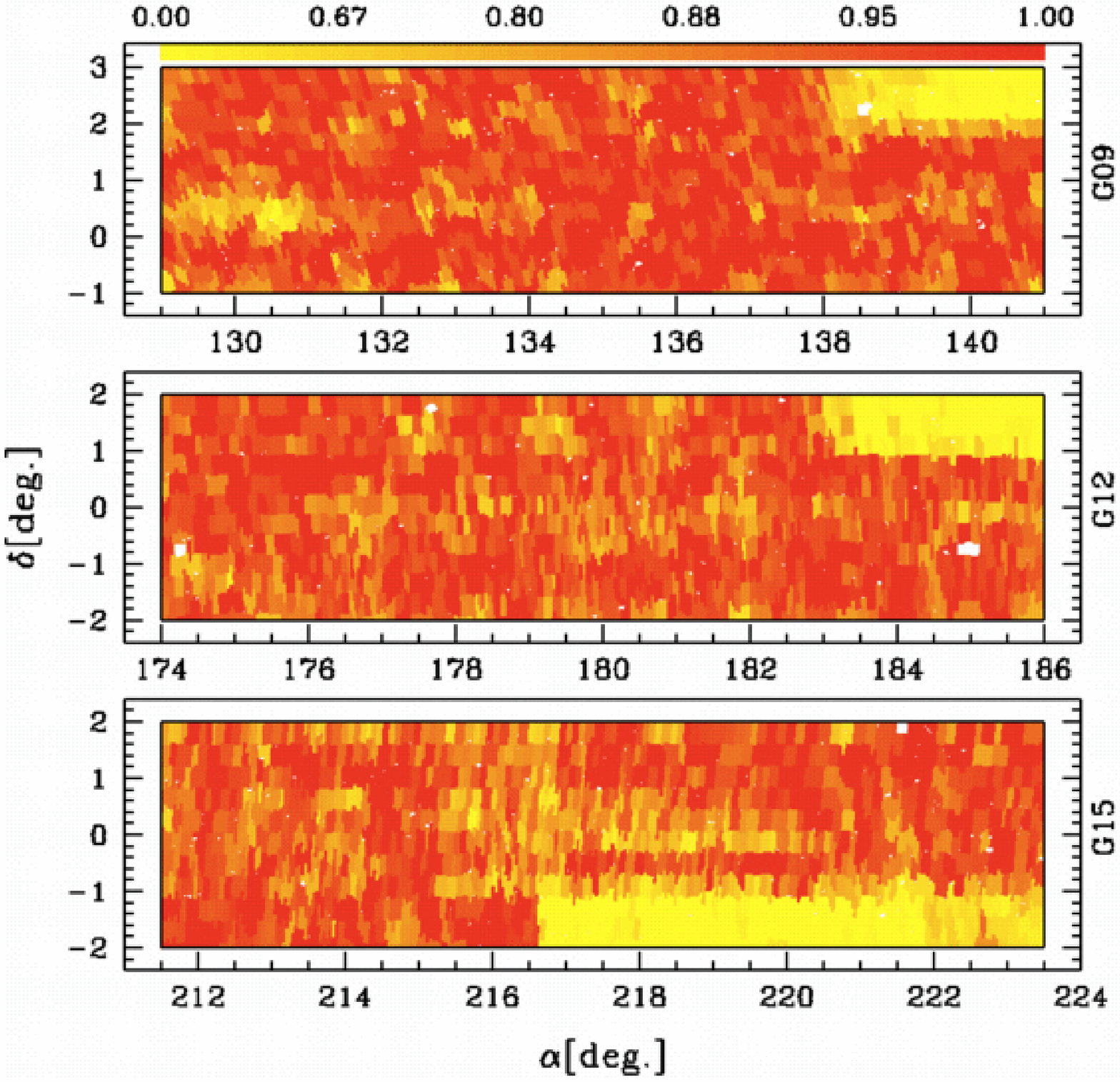,width=\columnwidth}}

\vspace{-12.0cm}

\hspace{8.0cm} {\psfig{file=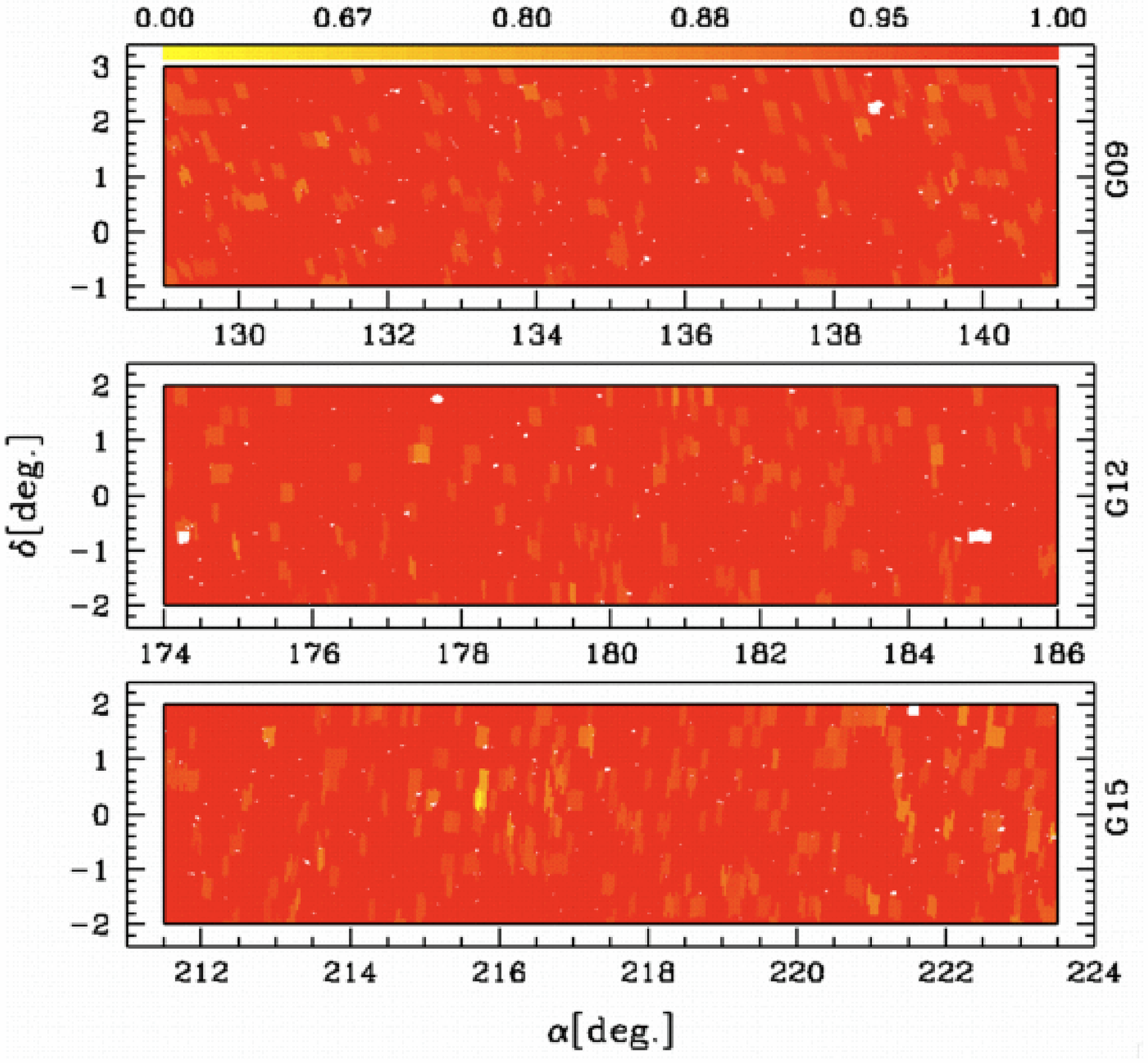,width=\columnwidth}}

\vspace{-2.0cm}

\caption{Spatial completeness masks to $r_{\rm pet} < 19.0$ mag after
  year 1 (left) and year 3 (right). The three GAMA regions are shown
  from top to bottom as G09, G12 and G15. The white regions indicate
  areas not sample either due to missing input catalogue data or
  bright stars. \label{fig:masks}}
\end{figure*}

\subsection{The imaging mask}
We are interested in knowing which regions of the GAMA areas have not
been properly covered by the SDSS imaging survey, or which should be
excluded owing to the presence of bright stars. For that reason, we
want to map out SDSS imaging areas containing any of the following
information: bleeding pixel, bright star, satellite trail or hole.

First we create, following the imaging mask information available
on the SDSS DR6 website\footnote{
{\tt http://www.sdss.org/dr6/products/images/use\_masks.html} and
{\tt http://www.sdss.org/dr6/algorithms/masks.html}},
the associated convex polygons, delimiting areas for which
imaging information is either not available or could be corrupted
(as in the case of bleeding pixels). We primarily use the $r$-band
imaging mask information, as GAMA is nominally an $r$-band selected
survey, but for completeness we construct all five SDSS
imaging masks for the GAMA areas.

Then we build an additional bright star mask based on
stars down to $V<12$ in the Tycho 2, Tycho 1 and Hipparcos
catalogues. For each star we define an exclusion radius $r$,
defined as:
\begin{eqnarray}
r & = & R_s/0.8 \qquad \qquad \mbox{for}~ 10 < V \le 12 \\
r & = & R_s/0.5 \qquad \qquad \mbox{for}~ V \le 10
\end{eqnarray}
where $R_s$ is the scattered-light radius, as estimated based on the
circular region over which the star flux per pixel is greater than
five times  the sky noise level. Further details on the bright star
exclusion mask are given in section 3.3 of Baldry et al.~(2010).

The final imaging mask function is then the union of the two separate
functions. For ease of use, the imaging masks have been pixelated
using an equal area projection.

\subsection{The $r_{\rm pet}$ redshift completeness mask}
Normally a simple way to define a redshift success rate would be
to make use of the geometry defined by the complete set of 2$^\circ$
fields that were used to tile the survey region for spectroscopic
observations. However, due to the tiling strategy adopted in the first
year, due to the preselection of observing the brightest targets
first, and due to the much higher number density of galaxies than
can be accommodated in a single 2dF field, this simple and
straightforward approach does not account well enough for the spatial
incompleteness of the survey. Therefore for year 1 data, we had to
develop different completeness masks each defined for a different
magnitude limit interval:

\begin{itemize}

\item For G09 and G15, there are two completeness masks each: one for
galaxies brighter than $r_{\rm pet}=19.0$ and one for
$19.0 \le r_{\rm pet} < 19.4$.

\item For G12, there are three completeness masks: the same two as for
G09 and G15, as well as one for $19.4 \le r_{\rm pet} < 19.8$.

\end{itemize}

Once the samples for which completeness masks are needed have been
defined, one just needs to provide a reasonable definition for the
redshift success rate. We choose to tessellate the GAMA regions with a
large number of sectors. We incorporate the relevant imaging mask at
this stage, by imposing that sectors do not cover any regions masked
by the imaging mask. Each sector contains between 15 and 50 galaxy
targets, is limited in extent (less than 24 arcmin) and in size (less
than 225 arcmin$^2$). These conditions are necessary to avoid
shot-noise dominated masks, and to guarantee that small scale
information is preserved as much as possible. We note that in the
current implementation these sectors are not uniquely
defined,\footnote{This could be achieved by making each target galaxy
  the centre of a subsector and then increasing the radius of all
  subsectors and create sectors by the merger of overlapping
  subsectors.}
but once specified any given position on the sky belongs to a unique
sector. For each sector, ${\bf \theta}$, we define the redshift
success rate, $R_{z}({\bf \theta})$, for a sample of galaxies within
specified magnitude limits, as the ratio of the number of galaxies for
which good quality redshifts have been obtained, $N_{z}({\bf \theta})$,
to the total number of objects contained in the tiling catalogue,
$N_{t}({\bf \theta})$.
The redshift completeness of a given sector, $R_{z}({\bf \theta})$,
should be clearly distinguished from the redshift completeness of a
given 2dF field, $c_{F}$, since multiple overlapping fields can
contribute to a single sector and $c_{F}$ is a measure of the quality of
the observing conditions for galaxies observed at the same time.

We present in Fig.~\ref{fig:masks} the completeness masks for all
three regions for GamaCoreDR1 (left) and GamaCoreMainSurvey (right) to
$r_{\rm pet}<19.0$. Completeness masks for all regions and different
selections are available to the team members and collaborators only at
this stage.

\section{GAMA Science ready catalogues and data release 1}
The combination of our five input catalogues, as outlined in
section~\ref{sec:build}, constitutes our core GAMA catalogue of $\sim
1$ million galaxies lying within the GAMA regions and extending to
approximately $r_{\rm pet} < 22.0$ mag. This catalogue has
inhomogeneous selection, is liable to be incomplete towards the
faint-end, along with significant noise in the photometry at the very
faint limit and spurious detections. We therefore extract from this
dataset three science-ready catalogues and one overflow
catalogue. These four catalogues along with the complete SWARP'ed
mosaics of the GAMA regions, associated spectra bundles, images,
S\'ersic profiles and a variety of data inspection tools (including a
MySQL access point) constitute our first data release and are now
available via: {\tt http://www.gama-survey.org/}

\subsection{{\sf GamaCoreMainSurvey} - Available 01/07/12}
The {\sf GamaCoreMainSurvey} is the GAMA team's principal science
catalogue and is constructed from {\sf GamaCore} by removing all
objects outside our Main Survey limits as defined in Baldry et
al.~(2010). These limits are: $r_{\rm pet} < 19.4$ mag in G09 \& G15
and $r_{\rm pet} < 19.8$ mag in G12 (114\,441 objects).  $K_{\rm Kron}
< 17.6$ with $r_{\rm model} < 20.5$ mag (61393 objects) or $z_{\rm
  Model} < 18.2$ with $r_{\rm model} < 20.5$ mag (55534 objects) and
are selected using {\sf SURVEY\_CLASS$ > 3$}. For a description of the
{\sf SURVEY\_CLASS} parameter see Table~\ref{surveyclass}). This
amounts to 119\,778 objects in total of which 101\,576 are new
redshifts provided by GAMA and 18202 pre-existing. Fig.~\ref{fig:comp}
shows the completeness in the $r$-band however similar plots can
trivially be constructed in $z$ or $K$. The spectroscopic completeness
(which includes objects not targeted) is 98.2 per cent in $r$, 99.3
per cent in $z$ and 98.6 per cent in $K$. The parameters contained in
this catalogue are listed in Table~\ref{tab:gdr1}. As this is our main
science catalogue we place an embargo on its public release until
$1^{st}$ July 2012 but are open to requests for collaboration sent to:
{\tt gama@gama-survey.org}

\begin{table*}
\caption{The \textsc{survey\_class} parameter given in {\sf GamaTiling}.
  Objects take the higher value if satisfying more than one
  criteria. All objects with \textsc{survey\_class} $\ge 2$ pass the
  standard star-galaxy separation.  See Baldry et al.\ (2010) for
  details. \label{surveyclass}}
\begin{tabular}{crcl} \hline
value & number & frac.\,$Q\ge3$ & criteria and notes \\ \hline
7  &  59756 & 0.995 & $r_{\rm pet}<19.0$ and $\Delta_{\rm sg} > 0.25$ (high priority in Year~1)\\
6  &  54685 & 0.968 & $r_{\rm pet}<19.4$ or $r_{\rm pet}<19.8$ in G12 ($\ge 6$ for $r$-limited Main Survey)\\
5  &   1386 & 0.703 & $z_{\rm model} < 18.2$ ($\ge 5$ for SDSS-mag limited Main Survey) \\
4  &   3951 & 0.879 & $K_{\rm AB} < 17.6$  ($\ge 4$ for Main Survey) \\
3  &  33134 & 0.343 & $19.4<r_{\rm pet}<19.8$ in G09 or G15 (F2 fillers) \\
2  &  14861 & 0.196 & $g_{\rm model}<20.6$ or $r_{\rm model}<19.8$ or $i_{\rm model}<19.4$ in G12 (F3 fillers) \\
1  &   1255 & 0.577 & radio selected targets (F1 fillers, Ching et al.\ in preparation) \\
0  &    822 & 0.394 & $2 \le$ \textsc{vis\_class} $\le 4$ or fails star-galaxy separation for $r_{\rm fib} < 17$ \\ \hline
\end{tabular}
\end{table*}

\subsection{{\sf GamaCoreDR1} --- Available 2010-06-25}
The {\sf GamaCoreDR1} is the current GAMA Data Release 1 catalogue and
is constructed from {\sf GamaCoreMainSurvey} by applying a strict
$r$-band selection of $r_{\rm pet} < 19.4$ mag in G09 \& G15 and
$r_{\rm pet} < 19.8$ mag in G12 (this is implemented by extracting
objects with {\sf SURVEY\_CLASS} $>5$.  While the final GAMA survey is
independently $r$, $z$ and $K$ selected, a strict $r$-band cut is
imposed because observations during year 1 were $r$-band limited only
with the additional cuts coming in in years 2 and 3. Future releases
involving year 2 and 3 data will therefore include $z$ and $K$ band
selected samples. Selecting on {\sf SURVEY\_CLASS} $>5$ yields a
catalogue of 114441 objects spread across the three GAMA
regions. Table~\ref{tab:gdr1} contains a list of the parameters being
released at this time. Redshift information (i.e., quality,
signal-to-noise etc.) are provided for all targets while the redshifts
are only provided for year 1 observations above $r_{\rm pet} = 19.0$
mag (52324 objects) and for a deeper narrow strip in G12 with
declination $\delta = \pm 0.5^{\circ}$ to $19.0 < r_{\rm pet} < 19.8$
mag (7533 objects). These limits are chosen as they represent our
year 1 spectroscopic targeting limits (see Figs~\ref{fig:comp},
\ref{fig:bias} \& \ref{fig:masks} left). The redshift release
constitutes approximately 50 per cent (59479 objects; 41902 GAMA;
17577 pre-existing), of the entire GAMA dataset (or 32.7 per cent of
the redshifts acquired by the GAMA team over the past three
years). This dataset provides a clean, well defined, $r$-band selected
sample suitable for scientific exploitation as shown by
Fig.~\ref{fig:comp} and Fig.~\ref{fig:bias}. Please note that
redshifts outside the specified flux limits or collected after year 1
are currently denoted as Z\_HELIO $= -2$. The redshift
quality and signal-to-noise information refers to the final best spectrum
held in {\sf GamaCoreMainSurvey} providing advance information as to
whether a reliable redshift exists in the larger database. Requests
for individual redshifts for specific objects can be directed to: {\tt
  gama@gama-survey.org}

\begin{figure}
\centerline{\psfig{file=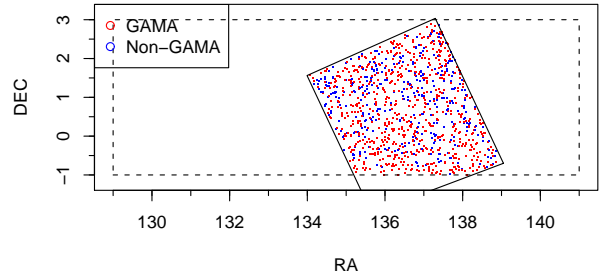,width=\columnwidth}}
\caption{The location of the H-ATLAS SV region overlaid on the
  GAMA 9hr region (dotted lines) and showing the location of
  pre-existing (blue) and new GAMA (red) redshifts within the common
  region. \label{fig:hatlassv}}
\end{figure}

\subsection{{\sf GamaCoreAtlasSV} --- Available 2010-06-25}
The {\sf GamaCoreAtlasSV} is the subset of objects that have both a
GAMA and H-ATLAS detection (see Smith et al. 2010 for details of
source matching). 1175 objects coexist between the two data sets in a
region defined by the geometric overlap between G09 and the H-ATLAS SV
region (see Fig.~\ref{fig:hatlassv}), with an approximate area of
$\sim16.7$ deg$^{2}$. Using the GamaCoreMainSurvey catalogue we find
$\sim15$ per cent of GAMA Main Survey targets have a H-ATLAS source
match. Of these matches 89 do not have any available redshift, 306
have SDSS redshifts, 792 have GAMA redshifts and 1 object has a
2SLAQ-QSO redshift. The GAMA redshifts comprise of 523 year 1
redshifts, 182 year 2 redshifts and 87 year 3 redshifts. There are no
magnitude cuts imposed for {\sf GamaCoreAtlasSV} and therefore 178 of
these redshifts derive from {\sf GamaCoreExtra}. Note that to extract
a homogeneously selected sample one must trim this catalogue to
$r_{pet} < 19.4$ mag. Table~\ref{tab:gsv} shows the parameters
included in this catalogue.

\begin{figure}
\centerline{\psfig{file=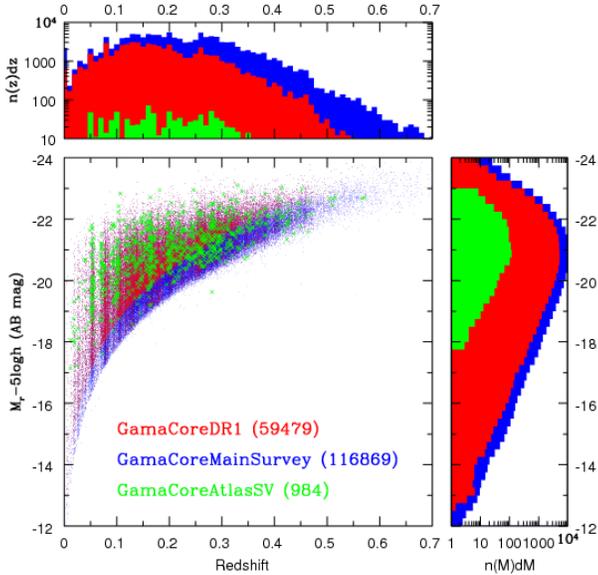,width=\columnwidth}}

\caption{(Main panel). The absolute magnitude versus redshift plane
  for the three science ready catalogues (as indicated). (left) the
  distribution collapsed in redshift, (upper) the distribution
  collapsed in absolute magnitude. \label{fig:core}}
\end{figure}

\subsection{{\sf GamaCoreExtraDR1} --- Available 2010-06-25}
The {\sf GamaCoreExtra} catalogue contains information on all objects
for which redshifts are known but lie outside the {\sf GamaCoreDR1}
selection limits. In effect {\sf GamaCoreExtra} serves as a redshift
dumping ground and when combined with {\sf GamaCoreMainSurvey} or {\sf
  GamaCoreDR1} provides a full record of all objects with known
redshifts in the three GAMA regions. This will include pre-existing
redshifts at faint magnitudes in the public domain (predominantly from
QSO and LRG surveys), along with filler objects targeted during the
GAMA campaign. The parameters included are identical to {\sf
  GamaCoreDR1} and therefore also described in
Table~\ref{tab:gdr1}. {\sf GamaCoreExtraDR1} is an identical copy of
{\sf GamaCoreExtra} except with the GAMA redshifts embargoed until a
later release. {\sf GamaCoreExtra} is highly inhomogeneous and
volatile as additional redshifts come to light and essentially forms
the starting point for any extension to the GAMA survey. Currently it
contains $\sim 16$k galaxies of which $\sim0.5$k are pre-existing and
$\sim15.5$k have been acquired during GAMA observations. Requests for
access to individual redshifts or redshifts for specified sub-regions
should be directed to: {\tt gama@gama-survey.org}

\subsection{Future plans}
The three science ready catalogues described in the previous sections
are shown in Fig.~\ref{fig:core} in the Absolute $r$-band magnitude
($M_r-5\log h$) versus redshift ($z$) plane (main panel), along with
the collapsed $N(z)$ and $N(M_r)$ distributions (upper and right
panels respectively). The absolute magnitudes are calculated using
$k$-corrections derived from the GAMA $u$ to $K$ photometry, and total
throughput (including atmosphere) temperature adjusted filter
transmission curves, and the {\sf KCORRECT} software package (Blanton
\& Roweis~2007). The figure is intended to provide a rough indication
of the parameter range being explored with $L^*$ galaxies routinely
detected to $z \approx 0.3$ and objects as faint as $M_r = -12$ mag
being recovered at $z>0.01$. In future papers we will explore more
comprehensively the completeness of the initial input catalogue ({\sf
  GamaTiling}) in the bivariate brightness distribution plane (BBD;
Hill et al.~2010, in prep.,) through a detailed BBD-analysis (see
Driver et al.~2005), as well as through comparisons to deeper and
higher spatial resolution data.

In the near term the GAMA team will be producing a number of extended
data products or data management units (DMUs), which include: spectral
line analysis via {\sf GANDALF}/ppxf (Sarzi et al., 2006; Cappellari
\& Emsellem 2004) and stellar population model fits (Maraston 2005;
Thomas, Maraston \& Bender 2003), bulge-disc decomposition via {\sf
  GALFIT3}, stellar masses and revised photometric-redshifts via {\sf
  INTEREST}, environmental markers, and group catalogues. These data
products will be outlined in more detail in our next data release
later in the year (Liske et al., 2010, in prep.). Within the next
3---5 years we expect to ingest additional datasets from GALEX,
HERSCHEL, WISE, GMRT, VISTA, VST, and ASKAP as well as expand our core
AAT programme to cover a larger area of $\sim 400$ sq deg. to $r_{\rm
  Sersic} < 19.8$ mag. In due course DMUs on multiwavelength SED and
HI modeling will provide dust mass/temperature measurements, and gas
and dynamical mass measurements.

\section{Summary}
The GAMA basic-data including GAMA and pre-GAMA spectra and redshifts,
GAMA SWARP processed images in $ugrizYJHK$, $u$ to $K$ matched aperture
photometry, corrections to total magnitudes are now available at: {\tt
  http://www.gama-survey.org/}. GAMA Data Release 1 ({\sf
  GamaCoreDR1}) data is publicly available via a downloads page
immediately, and the full catalogue ({\sf GamaCoreMainSurvey}) is
available for use by the GAMA team and collaborators. In addition we
make publicly available a catalogue ({\sf GamaCoreAtlasSV}) including
year-2 data for the H-Atlas Science Verification region, which
form the basis for a number of H-Atlas Science Verification
papers.

Data inspections tools, including a MySQL tool and direct data
inspection toolkit, are provided and will be developed further in due
course. The sum of year 1, 2, and 3 data constitutes GAMA Phase I,
which is now complete having utilised archival data from the Sloan
Digital Sky Survey, the UK Infrared Deep Sky Survey, and the NASA
Extragalactic Database and an initial allocation of 66 nights of
Anglo-Australian Telescope time. Value added data products will be
released from January 1st 2011 and regularly thereafter. GAMA Phase II
is now underway and aims to add in additional area coverage as well
as to extend the depth of the survey. The community is encouraged and
invited to contact the team for early access to {\sf
  GamaCoreMainSurvey}. The contents of this catalogue are described
in Table~\ref{tab:gdr1}.

\section*{acknowledgments}
First and foremost we would like to thank the staff of the former
Anglo-Australian Observatory (now Australian Astronomical Observatory),
for the the smooth running of the telescope and AAOmega during these
observations and the funding support we have received from the STFC
and the ARC.

Funding for the SDSS and SDSS-II has been provided by the Alfred
P. Sloan Foundation, the Participating Institutions, the National
Science Foundation, the U.S. Department of Energy, the National
Aeronautics and Space Administration, the Japanese Monbukagakusho, the
Max Planck Society, and the Higher Education Funding Council for
England. The SDSS Web Site is {\tt http://www.sdss.org/}.  The SDSS is
managed by the Astrophysical Research Consortium for the Participating
Institutions. The Participating Institutions are the American Museum
of Natural History, Astrophysical Institute Potsdam, University of
Basel, University of Cambridge, Case Western Reserve University,
University of Chicago, Drexel University, Fermilab, the Institute for
Advanced Study, the Japan Participation Group, Johns Hopkins
University, the Joint Institute for Nuclear Astrophysics, the Kavli
Institute for Particle Astrophysics and Cosmology, the Korean
Scientist Group, the Chinese Academy of Sciences (LAMOST), Los Alamos
National Laboratory, the Max-Planck-Institute for Astronomy (MPIA),
the Max-Planck-Institute for Astrophysics (MPA), New Mexico State
University, Ohio State University, University of Pittsburgh,
University of Portsmouth, Princeton University, the United States
Naval Observatory, and the University of Washington.

The UKIDSS project is defined in Lawrence et al. (2007). UKIDSS uses
the UKIRT Wide Field Camera (WFCAM; Casali et al, 2007). The
photometric system is described in Hewett et al. (2006), and the
calibration is described in Hodgkin et al. (2009). The pipeline
processing and science archive are described in Irwin et al. (2009, in
prep) and Hambly et al. (2008). We have used data from the 4th data
release.

This research has made use of the NASA/IPAC Extragalactic Database
(NED), which is operated by the Jet Propulsion Laboratory, California
Institute of Technology, under contract with the National Aeronautics
and Space Administration.

The Millennium Galaxy Catalogue consists of imaging data from the
Isaac Newton Telescope and spectroscopic data from the Anglo
Australian Telescope, the ANU 2.3m, the ESO New Technology Telescope,
the Telescopio Nazionale Galileo and the Gemini North Telescope. The
survey has been supported through grants from the Particle Physics and
Astronomy Research Council (UK) and the Australian Research Council
(AUS). The data and data products are publicly available from
{\tt http://www.eso.org/{\tt\char'176}jliske/mgc/} or on request from J. Liske or
S.P. Driver.

Finally we would like to acknowledge the use of Topcat, Stilts,
SExtractor, SWARP, PSFex, GALFIT3 and ALADIN astronomical software
packages and the r data analysis package.

\appendix

\section{GAMA catalogues}
The four principle GAMA catalogues formed at this time are: {\sf
  GamaTiling}, {\sf GamaCoreMainSurvey}, {\sf GamaCoreDR1} and {\sf
  GamaCore AtlasSv}. The parameters contained in these catalogues and
their explanations are showing in Tables.~\ref{tab:gt}---\ref{tab:gsv}
respectively.

\begin{table*}
\caption{Parameters held in {\sf GamaTiling} \label{tab:gt}}
\begin{tabular}{cccp{11.0cm}} \hline 
Column & Parameter & Units & Definition \\ \hline 
1  & GAMA\_ID & N/A & Unique six digit GAMA identifier linked to GAMAs original SDSS DR6 input cat \\ 
2  & SDSS\_ID & N/A & Sloan Identifier from SDSS DR6 \\ 
3  & RA\_J2000 & Degrees & RA taken from SDSS DR6 \\ 
4  & DEC\_J2000 & Degrees & Declination taken from SDSS DR6 \\ 
5  & r\_FIBREMAG & AB mag & Fibre magnitude \\
6  & r\_PETRO & AB mag & Extinction corrected $r$-band Petrosian flux derived from SDSS DR6 \\ 
7  & u\_MODEL & AB mag & Extinction corrected model $u$ mag from SDSS DR6 \\
8  & g\_MODEL & AB mag & Extinction corrected model $g$ mag from SDSS DR6 \\
9  & r\_MODEL & AB mag & Extinction corrected model $r$ mag from SDSS DR6 \\
10 & i\_MODEL & AB mag & Extinction corrected model $i$ mag from SDSS DR6 \\
11 & z\_MODEL & AB mag & Extinction corrected model $z$ mag from SDSS DR6 \\
12 & NUM\_GAMA\_SPEC & N/A & Number of spectra available for this object \\
13 & r\_SB & AB mag arcsec$^{-2}$ & SDSS Petro half-light $r$-band surface brightness \\
14 & SG\_SEP & mag & SDSS r\_psf-r\_model star-galaxy separation parameter \\
15 & SG\_SEP\_JK & mag & GAMA AUTO J-K - f\_locus star-galaxy separation parameter \\
16 & K\_KRON\_SELECTION & AB mag & Extinction corrected K magnitude used for selection (Old)  \\
17 & TARGET\_FLAGS & bit & bitwise target criteria (see gama website) \\
18 & SURVEY\_CLASS & N/A & Survey class (see Table~\ref{surveyclass})  \\
19 & PRIORITY\_CLASS & N/A & Priority class (see gama website) \\
20 & NEIGHBOUR\_CLASS & N/A & Number of Main Survey neighbours within 40 arcseconds (see Fig.~\ref{fig:pairs})\\
21 & MASK\_IC\_10 & N/A & Mask value 0.0 to 1.0 around $V < 10$ mag stars (see Baldry et al.~2010)\\
22 & MASK\_IC\_12 & N/A & Mask value 0.0 to 1.0 around $V < 12$ mag stars (see Baldry et al.~2010)\\
23 & VIS\_CLASS & N/A & Visual classification (see Baldry et al.~2010) \\
24 & VIS\_CLASS\_USER & N/A & Initials of visual classifier for VIS\_CLASS \\ \hline
\end{tabular}
\end{table*}

\pagebreak

\begin{table*}
\caption{Parameters held in {\sf GamaCoreMainSurvey}, {\sf GamaCoreDR1}, {\sf GamaCoreExtraDR1}. \label{tab:gdr1}}
\begin{tabular}{cccp{11.0cm}} \hline
Column & Parameter & Units & Definition \\ \hline 
1 & GAMA\_IAU\_ID & N/A & Unique IAU identifier \\ 
2 & GAMA\_ID & N/A & Unique six digit GAMA identifier linked to GAMAs original SDSS DR6 input cat \\ 
3 & SDSS\_ID & N/A & Sloan Identifier from SDSS DR6 \\ 
4 & RA\_J2000 & Degrees & RA taken from SDSS DR6 \\ 
5 & DEC\_J2000 & Degrees & Declination taken from SDSS DR6 \\ 
6 & r\_PETRO & AB mag & Extinction corrected $r$-band Petrosian flux derived from SDSS DR6 \\ 
7 & Z\_HELIO & N/A & Heliocentric redshift ($-2$ = embargoed, $=-0.9$ if nQ=1)\\ 
8 & Z\_QUALITY & N/A & Confidence on redshift (Definite, nQ=4, Reliable nQ=3, Uncertain nQ=2, Unknown nQ=1, Not observed nQ=99) \\ 
9 & Z\_SOURCE & N/A & Origin of redshift 1=SDSS DR6, 2=2dFGRS, 3=MGC, 4=2SLAQ-LRG, 5=GAMA, 6=6dFGS, 7=UZC, 8=2QZ, 9=2SLAQ-QSO, 10=NED \\ 
10 & Z\_DATE & N/A & Date of observation if z\_SOURCE=5 (i.e., GAMA), otherwise 99\\ 
11 & Z\_SN & N/A & Mean s/n of spectrum if z\_SOURCE=5 (i.e., GAMA), otherwise 99 \\ 
12 & Z\_ID & N/A & Filename of best available spectrum. \\ 
13 & PHOT\_SOURCE & N/A & Photometric source (rd = $r$-band defined, sd = self-defined, see Hill et al. 2010a)\\ 
14 & u\_KRON & AB mag & Extinction $u$-band Kron magnitude derived from $r$-band aperture \\ 
15 & g\_KRON & AB mag & Extinction $g$-band Kron magnitude derived from $r$-band aperture \\ 
16 & r\_KRON & AB mag & Extinction $r$-band Kron magnitude derived from $r$-band aperture \\ 
17 & i\_KRON & AB mag & Extinction $i$-band Kron magnitude derived from $r$-band aperture \\ 
18 & z\_KRON & AB mag & Extinction $z$-band Kron magnitude derived from $r$-band aperture \\ 
19 & Y\_KRON & AB mag & Extinction $Y$-band Kron magnitude derived from $r$-band aperture \\ 
20 & J\_KRON & AB mag & Extinction $J$-band Kron magnitude derived from $r$-band aperture \\ 
21 & H\_KRON & AB mag & Extinction $H$-band Kron magnitude derived from $r$-band aperture \\ 
22 & K\_KRON & AB mag & Extinction $K$-band Kron magnitude derived from $r$-band aperture \\ 
23 & u\_KRON\_ERR & AB mag & $u$-band Kron magnitude error \\ 
24 & g\_KRON\_ERR & AB mag & $g$-band Kron magnitude error \\ 
25 & r\_KRON\_ERR & AB mag & $r$-band Kron magnitude error \\ 
26 & i\_KRON\_ERR & AB mag & $i$-band Kron magnitude error \\ 
27 & z\_KRON\_ERR & AB mag & $z$-band Kron magnitude error \\ 
28 & Y\_KRON\_ERR & AB mag & $Y$-band Kron magnitude error \\ 
29 & J\_KRON\_ERR & AB mag & $J$-band Kron magnitude error \\ 
30 & H\_KRON\_ERR & AB mag & $H$-band Kron magnitude error \\ 
31 & K\_KRON\_ERR & AB mag & $K$-band Kron magnitude error \\ 
32 & EXTINCTION\_r & AB mag & Galactic magnitude extinction in $r$-band \\
33 & r\_SERS\_MAG\_10RE & AB mag & $r$-band S\'ersic magnitude truncated at 10 half light radii \\
\end{tabular}
\end{table*}

\pagebreak

\begin{table*}
\caption{Parameters held in {\sf GamaCoreAtlasSV} \label{tab:gsv}}
\begin{tabular}{cccp{11.0cm}} \hline
Column & Parameter & Units & Definition \\ \hline 
1 & GAMA\_IAU\_ID & N/A & IAU certified GAMA ID \\ 
2 & GAMA\_ID & N/A & Unique six digit GAMA identifier linked to GAMAs original SDSS DR6 input cat \\ 
3 & SDSS\_ID & N/A & Sloan Identifier from SDSS DR6 \\ 
4 & RA\_J2000 & Degrees & RA taken from SDSS DR6 \\ 
5 & DEC\_J2000 & Degrees & Declination taken from SDSS DR6 \\ 
6 & r\_PETRO & AB mag & Extinction corrected $r$-band Petrosian flux derived from SDSS DR6 \\ 
7 & Z\_HELIO & N/A & Heliocentric redshift ($-0.9$ if nQ=1) \\ 
8 & Z\_QUALITY & N/A & Confidence on redshift (Definite, nQ=4, Reliable nQ=3, Uncertain nQ=2, Unknown nQ=1, Not observed nQ=99) \\ 
9 & Z\_SOURCE & N/A & Origin of redshift 1=SDSS DR6, 2=2dFGRS, 3=MGC, 4=2SLAQ-LRG, 5=GAMA, 6=6dFGS, 7=UZC, 8=2QZ, 9=2SLAQ-QSO, 10=NED \\ 
10 & Z\_DATE & N/A & Date of observation if z\_SOURCE=5 (i.e., GAMA), otherwise 99\\ 
11 & Z\_SN & N/A & Mean s/n of spectrum if z\_SOURCE=5 (i.e., GAMA), otherwise 99 \\ 
12 & Z\_ID & N/A & Filename of best available spectrum. \\ 
13 & PHOT\_SOURCE & N/A & Photometric source (rd = $r$-band defined, sd = self-defined, see Hill et al. 2010a)\\ 
14 & u\_KRON & AB mag & Extinction $u$-band Kron magnitude derived from $r$-band aperture \\ 
15 & g\_KRON & AB mag & Extinction $g$-band Kron magnitude derived from $r$-band aperture \\ 
16 & r\_KRON & AB mag & Extinction $r$-band Kron magnitude derived from $r$-band aperture \\ 
17 & i\_KRON & AB mag & Extinction $i$-band Kron magnitude derived from $r$-band aperture \\ 
18 & z\_KRON & AB mag & Extinction $z$-band Kron magnitude derived from $r$-band aperture \\ 
19 & Y\_KRON & AB mag & Extinction $Y$-band Kron magnitude derived from $r$-band aperture \\ 
20 & J\_KRON & AB mag & Extinction $J$-band Kron magnitude derived from $r$-band aperture \\ 
21 & H\_KRON & AB mag & Extinction $H$-band Kron magnitude derived from $r$-band aperture \\ 
22 & K\_KRON & AB mag & Extinction $K$-band Kron magnitude derived from $r$-band aperture \\ 
23 & u\_KRON\_ERR & AB mag & $u$-band Kron magnitude error \\ 
24 & g\_KRON\_ERR & AB mag & $g$-band Kron magnitude error \\ 
25 & r\_KRON\_ERR & AB mag & $r$-band Kron magnitude error \\ 
26 & i\_KRON\_ERR & AB mag & $i$-band Kron magnitude error \\ 
27 & z\_KRON\_ERR & AB mag & $z$-band Kron magnitude error \\ 
28 & Y\_KRON\_ERR & AB mag & $Y$-band Kron magnitude error \\ 
29 & J\_KRON\_ERR & AB mag & $J$-band Kron magnitude error \\ 
30 & H\_KRON\_ERR & AB mag & $H$-band Kron magnitude error \\ 
31 & K\_KRON\_ERR & AB mag & $K$-band Kron magnitude error \\ 
32 & EXTINCTION\_r & AB mag & Galactic magnitude extinction in $r$-band \\
33 & r\_SERS\_MAG\_10RE & AB mag & $r$-band S\'ersic magnitude truncated at 10 half light radii \\
34 & HATLAS\_IAU\_ID & N/A & H-ATLAS ID as specified by Smith et al. (2010) \\ \hline
\end{tabular}
\end{table*}

\pagebreak

\section*{References}

\reference Adelman-McCarthy J.K., et al., 2008, ApJS, 175, 297

\reference Allen P.D., Driver S.P., Graham A.W., Cameron E., Liske J., De Propris R., 2006, MNRAS, 371, 2

\reference Abazajian K.N., et al., 2009, ApJS 182, 543

\reference Baldry, I.K., et al., 2010, MNRAS, 404, 86

\reference Baldry I.K., Balogh M.L., Bower R.G., Glazebrook K., Nichol R.C., Bamford S.P., Budavari T., 2006, MNRAS, 373, 469 

\reference Baugh C.M., et al., 2004, MNRAS, 351, 44

\reference Belokurov V., et al., 2006, ApJ, 642, 137

\reference Bell E.F., McIntosh D.H., Katz N., Weinberg M.D., 2003, ApJS, 149, 289

\reference Bell E.F., Phleps S., Somerville, R.S., Wolf, C., Borch A., Meisenheimer, K., 2006, ApJ, 652, 270

\reference Bertin E., Arnouts S., 1996, A\&AS, 117, 393

\reference Bertin E., Mellier Y., Radovich M, Missonnier G, Didelon P,
Morin B, 2002, in ASP Conf. Proc on Astronomical Data Analysis
Software and Systems (Eds. David A. Bohlender, Daniel Durand \& Thomas
H. Handley), 281, 228

\reference Blanton M.R., Roweis, S., 2007, AJ, 113, 734

\reference Bower R.G., Benson A.J., Malbon R, Helley J.C., Frenk C.S., Baugh C.M., Cole S.G., Lacey C.G., 2006, MNRAS, 370, 645

\reference Canon R., et al., 2006, MNRAS, 372, 425

\reference Cameron E., Driver S.P., 2007, MNRAS, 377, 523 

\reference Cappellari M., Emsellem E., 2004, PASP, 116, 138

\reference Choi Y-Y, Park Changbom, Vogeley M.S., 2007, ApJ, 638, 87

\reference Cole S., et al., 2005, MNRAS, 362, 505

\reference Colless M., et al., 2001, MNRAS, 328, 1039

\reference Colless M., et al., 2003, arXiv:astro-ph/0306581

\reference Collister A.A., Lahav O., 2004, PASP, 116, 345

\reference Cook M., Barausse E., Evoli C., Lapi A., Granato G.L., 2010, MNRAS, 402, 2113

\reference Croom S., Saunders W., Heald R., 2004, AAO Newsletter, 106, 12

\reference Croom S.M., Smith R.J., Boyle B.J., Shanks T., Miller L., Outram P.J., Loaring N.S., 2004, MNRAS, 349, 1397

\reference Croom S.M., et al., 2009, MNRAS, 392, 19

\reference De Lucia G., Springer V., White S.D.M., Croton D., Kauffmann G., 2006, MNRAS, 366, 499 

\reference De Propris R, Conselice C.J., Liske J., Driver S.P., Patton D.R., Graham A.W., Allen P.D., 2007, ApJ, 666, 212

\reference Driver, S.P., Liske J., Cross N.J.G., De Propris R., Allen P.D., 2005, MNRAS, 360, 81

\reference Driver S.P., et al., 2006, MNRAS, 368, 414

\reference Driver S.P., Popescu C.C., Tuffs R.J., Liske J., Graham A.W., Allen P.D., De Propris, R, 2007, MNRAS, 379, 1022

\reference Driver S.P., Popescu C.C., Tuffs R.J., Graham A.W., Liske J., Baldry I., 2008, ApJ, 678, 101

\reference Driver S.P., et al., 2009, A\&G, 50, 12 

\reference Driver S.P., Robotham A.S.G., 2010, MNRAS, in press (arXiv:1005.2538)

\reference Eales, S., et al., 2010, PASP, 122, 499

\reference Eke V.R., et al., 2004, MNRAS, 348, 866

\reference Falco E.E., et al., 1999, PASP, 111, 438

\reference Gadotti, D.A., 2009, MNRAS, 393, 1531

\reference Graham A.W., Driver S.P., 2005, PASA, 22, 118

\reference Guzzo L. et al., 2008, Nature, 451, 541

\reference Heavens, A., Panter B., Jimenez R., Dunlop J., 2004, Nature, 428, 625

\reference Hill, D., et al., 2010a, MNRAS, submitted

\reference Hill, D., Driver S.P., Cameron E.C., Cross N.J.G., Liske J., 2010b, MNRAS, 404, 1215

\reference Hopkins A.M., McClure-Griffiths N.A., Gaensler B.M., 2008, ApJ, 682, L13

\reference Hopkins P., Hernquist L, Cox T.J., Di Matteo T., Robertson B., Springel V., 2006, ApJS, 163, 1

\reference Ilbert, O., et al., 2009, ApJ 690, 1236

\reference Johnston, S., et al., 2007, PASA, 24, 174

\reference Jones D.H., et al., 2004, MNRAS, 355, 747

\reference Jones D.H., et al., 2009, MNRAS, 399, 683

\reference Kirkpatrick J. D., et al., 1999, ApJ, 519, 802 

\reference Kron R.G., 1980, ApJS, 43, 305

\reference Lah, P., et al., 2009, MNRAS, 399, 1447

\reference Lawrence A., et al., 2007, MNRAS, 379, 1599

\reference Lewis I.J., et al., 2002, MNRAS, 333, 279

\reference Lilly S.J., et al., 2007, ApJS, 172, 70

\reference Lupton R., Blanton M.R., Fekete G., Hogg D.W., O'Mullane W., Szalay A., Wherry N., 2004, PASP, 116, 133

\reference Maraston C., 2005, MNRAS, 362, 799

\reference Martin D.C., et al., 2005, ApJ, 619, 1

\reference Masters K.L., et al., 2010, 404, 792

\reference Merritt D., Graham A.W., Moore, B., Diemand J., Terzic, B., 2006, AJ, 132, 2685

\reference Meyer, M., et al., 2004, MNRAS, 350, 1195

\reference Nichol R.C., et al., 2006, MNRAS, 368, 1507

\reference Nieto-Santisteban M.A., Szalay A.S., Gray J., 2004, in ASP
Conf. Proc on Astronomical Data Analysis Software Systems
(Eds. Francois Ochsenbein, Mark G Allen \& Daniel Egret), 314, 666

\reference Norberg P., et al., 2001, MNRAS, 328, 64

\reference Novak G.S., Faber S.M., Dekel A., 2006, ApJ, 637, 96


\reference Padmanabhan N., et al., 2008, ApJ 674, 1217

\reference Peng, C.Y., Ho L.C., Impey C.D., Rix H-W., 2010, AJ, 139, 2097

\reference Percival W.J., et al., 2001, MNRAS, 327, 1297

\reference Petrosian V., 1976, ApJ, 209, 1

\reference Pohlen M., Trujillo I., 2006, A\&A, 454, 759

\reference Popescu C.C., Misiriotis A., Kylafis N.D., Tuffs R.J.,
Fischera J., 2000, A\&A, 362, 138

\reference Robotham A.S.G., et al., 2010, PASA, 27, 76

\reference Sarzi M., et al., 2006, MNRAS, 366, 1151

\reference S\'ersic J.L., 1963, BAAA, 6, 41

\reference S\'ersic J.L., 1968, in Observatorio Astronomico, (Publ: ESO)

\reference Shao Z., Xiao Q., Shen S., Mo H.J., Xia X. Deng Z., ApJ, 659, 1159

\reference Sharp, R., et al., 2006, in SPIE proceedings on Ground-based
and Airborne Instrumentation for Astronomy (Eds. Ian S. McLean,
Masanori Iye), 6269, 62690

\reference Sharp R., Parkinson H., 2010, MNRAS, in press (arXiv:1007.0648)

\reference Skrutskie M.F., et al., 2006, AJ, 131, 1163

\reference Smith, D.J.B., et al., 2010, MNRAS, submitted (arXiv:1007.5260)

\reference Spergel D., et al., 2003, ApJS, 148, 175

\reference Thomas D., Maraston C., Bender R., 2003, MNRAS, 339, 897

\reference Trimble V., Ceja, J.A., 2010, AN, 331, 338

\reference York D.G., et al., 2000, AJ, 120, 1579

\reference Watson F., Colless M.C., 2010, A \& G, 51, 3.16

\reference Wild, V., Hewett P.C., 2005, MNRAS, 358, 1083

\reference Willman B., et al., 2005, ApJ, 626, 85

\label{lastpage}

\end{document}